\documentclass[article]{solarphysics}
\usepackage[optionalrh,solaromanenum,natbib]{spr-sola-addons} 
\usepackage{graphicx}                    
\usepackage{color}                       
\usepackage{url}                         

\usepackage{rotating}
\usepackage{longtable}
\usepackage{lscape}
\usepackage{epstopdf}
\usepackage{xcolor,colortbl}
\usepackage{float}
\usepackage{hyperref}
\usepackage{array}
\usepackage{dcolumn}
\usepackage{color}
 \newcommand{\changes}[1]{#1}                         

\ifx \doiurl    \undefined \def \doiurl#1{\href{http://dx.doi.org/#1}{\textsf{DOI}}}\fi
\ifx \adsurl    \undefined \def \adsurl#1{\href{http://adsabs.harvard.edu/abs/#1}{\textsf{ADS}}}\fi
\ifx \arxivurl  \undefined \def \arxivurl#1{\href{http://arxiv.org/abs/#1}{\textsf{arXiv}}}\fi

\runningauthor{M.S. Freed {\it et al.}}
\runningtitle{Three-Year Global Survey of Coronal Null Points}

\begin{document}

\begin{article}

\begin{opening}

\title{Three-Year Global Survey of Coronal Null Points from Potential-Field-Source-Surface (PFSS) Modeling and Solar Dynamics Observatory (SDO) Observations}

%
\author{M.S.~\surname{Freed}$^{1}$\sep
       D.W.~ \surname{Longcope}$^{1}$\sep
        D.E.~\surname{McKenzie}$^{1}$      
       }
%

%
  \institute{$^{1}$Department of Physics, Montana State University,\\
  			 Bozeman, MT 59717, USA \\
                     email: \href{mailto:mfreed@physics.montana.edu}{mfreed@physics.montana.edu}
           }

\begin{abstract}
This article compiles and examines a comprehensive coronal magnetic-null-point survey created by potential-field-source-surface (PFSS) modeling and {\it Solar Dynamics Observatory/Atmospheric Imaging Assembly} (SDO/AIA) observations. The locations of 582 potential magnetic null points in the corona were predicted from the PFSS model between Carrington Rotations (CR) 2098 \changes{(June 2010)} and 2139 \changes{(July 2013)}. These locations were manually inspected, using contrast-enhanced SDO/AIA images in 171\,{\AA} at the east and west solar limb, for structures associated with nulls. A Kolmogorov--Smirnov (K--S) test showed a statistically significant difference between observed and predicted latitudinal distributions of null points. This finding is explored further to show that the observability of null points could be affected by the Sun's asymmetric hemisphere activity. Additional K--S tests show no effect on observability related to eigenvalues associated with the fan and spine structure surrounding null points or to the orientation of spine. We find that approximately 31\,\% of nulls obtained from the PFSS model were observed in SDO/AIA images at one of the solar limbs. An observed null on the east solar limb had a 51.6\,\% chance of being observed on the west solar limb. Predicted null points going back to CR 1893 \changes{(March 1995)} were also used for comparing radial and latitudinal distributions of nulls to previous work and to test for correlation of solar activity to the number of predicted nulls.

\end{abstract}

%

\end{opening}

%

\section{Introduction}
     \label{S-Introduction}
     From their conception, magnetic reconnection models have attached special significance to those locations in the corona where the magnetic field vanishes \citep{Sweet58,Dungey58}. Evidence of such null points were reported in observations taken by {\it Yohkoh/Soft X-ray Telescope} (SXT) and {\it SOHO/Extreme ultraviolet Imaging Telescope} (EIT) \citep{Filippov99}, which indicated the presence of a ``saddle" structure above AR\,8113. Observations from the first ``Bastille Day'' flare (SOL:1998-07-14T12:55:00L216C113) in 1998 also showed the role that null points can play in solar flares \citep{Aulanier00}, while \citet{Demoulin94} investigated how important these features are for solar flares in general. Also, \citet{Barnes07} used Imaging Vector Magnetograph data to show that active regions with null points have a greater chance of hosting a coronal mass ejection than those without a null. Other observations have shown the role that null points can potentially have in explaining coronal jets, flare ribbons, and coronal bright points \citep{Moreno08,Masson09,Zhang12}. Null points are also important in understanding magnetic reconnection in the Earth's magnetosphere from interactions with the solar wind \citep{Dorelli07,Xiao07}.\par
      Initial theoretical work centered on magnetic reconnection in two dimensions with null points called X-type nulls. However, advances in computing allowed for more sophisticated 3D behavior to be explored by \citet{Galsgaard97}, which was preceded by extensive analytical work on spine and fan reconnection by \citet{Priest96}. A considerable amount of work has also gone into identifying the magnetic skeleton associated with coronal fields, which includes locating null points \citep{Priest97,Longcope02,Titov11,Titov12,Platten14}. There is additional theoretical interest in understanding how flux can transfer between regions of open and closed fields surrounding a coronal null point (hereafter CNP)\citep{Pontin13} and how external disturbances such as magnetoacoustic waves behave near a null \citep{Galsgaard03}.\par
      Null points can occur in any magnetic field \changes{associated with astrophysical phenomena},  and the number of null points will depend on the spatial structure within that field.  \citet{Albright99} showed in general that the density of null points in a random, statistically homogeneous field depends on the distribution of spatial scales of that field.    Null points that can initiate magnetic reconnection in the solar corona will be those that occur in equilibrium fields. Toward that end, a number of authors have computed the distribution of null points in different kinds of potential fields anchored to prescribed photospheric fields.   \citet{Longcope03} considered a potential field anchored to a random and statistically homogeneous photospheric field.  \citet{Longcope09} and \citet{Longcope09b} applied this result to the quiet Sun using observations from {\it Solar and Heliospheric Observatory/Michelson Doppler Imager} (SOHO/MDI) and {\it Hinode/Narrowband filter imager} (NFI).  They found that the number of null points fell off with, approximately, the inverse third power of the height.\par
   At the largest scales, the coronal field cannot be considered homogeneous, so a particular photospheric field must be considered.  The largest-scale extrapolations are the potential-field-source-surface (PFSS) models \citep{Altschuler69,Schatten69}. These model fields can manifest magnetic null points that are sometimes observed to relate to coronal activity \citep{Savage10,Masson14}.  Several studies have found the distribution of magnetic null points in PFSS model fields \citep{Cook09,Platten14}.  The PFSS is, however, a theoretical model field known to depart from the field of the real solar corona \citep{Lowder14}.  One measure of the fidelity of the model could be obtained by comparing the number of null points predicted by the model to the number observed in coronal observations.  The present work is intended to make this comparison.\par
  The generic magnetogram configuration of a CNP consists of an embedded bipolar active region underneath the CNP site \citep{Antiochos98}. This same configuration has been used for numerical models of coronal jets which show similar structure \citep{Shibata92,Yokoyama96,Pariat09}. The primary features associated with CNPs are the fan that forms a separatrix surface and the spine as illustrated in Figure 4 of \cite{Longcope05}. The separatrix surface acts as a divider between field lines with different connectivity. A CNP is classified as positive if its fan consists of field lines diverging from the null, and \changes{negative if they are converging to the null}.  The spine consists of two field lines that originate or terminate at the null point. A comprehensive explanation of CNP structures and how to find them has been given by \citet{Greene88,Greene92}, \citet{Parnell96}, and \citet{Haynes07,Haynes10}. \par
     In Section 2, we describe how the CNPs were located from the PFSS model and then how SDO/AIA images were compiled to create a CNP catalog. Some general statistics from this survey and examples of the information available in the CNP catalog can be found in Section 3. The effect of null-point parameters on observability, properties of predicted nulls, and a follow-up investigation with USAF/NOAA sunspot data for determining hemispheric asymmetry are all discussed in Section 4. A summary of findings appears in Section 5.\par
     
         \begin{figure} [h]
 \centerline{\includegraphics[angle=-90,width=0.9\textwidth,clip=]{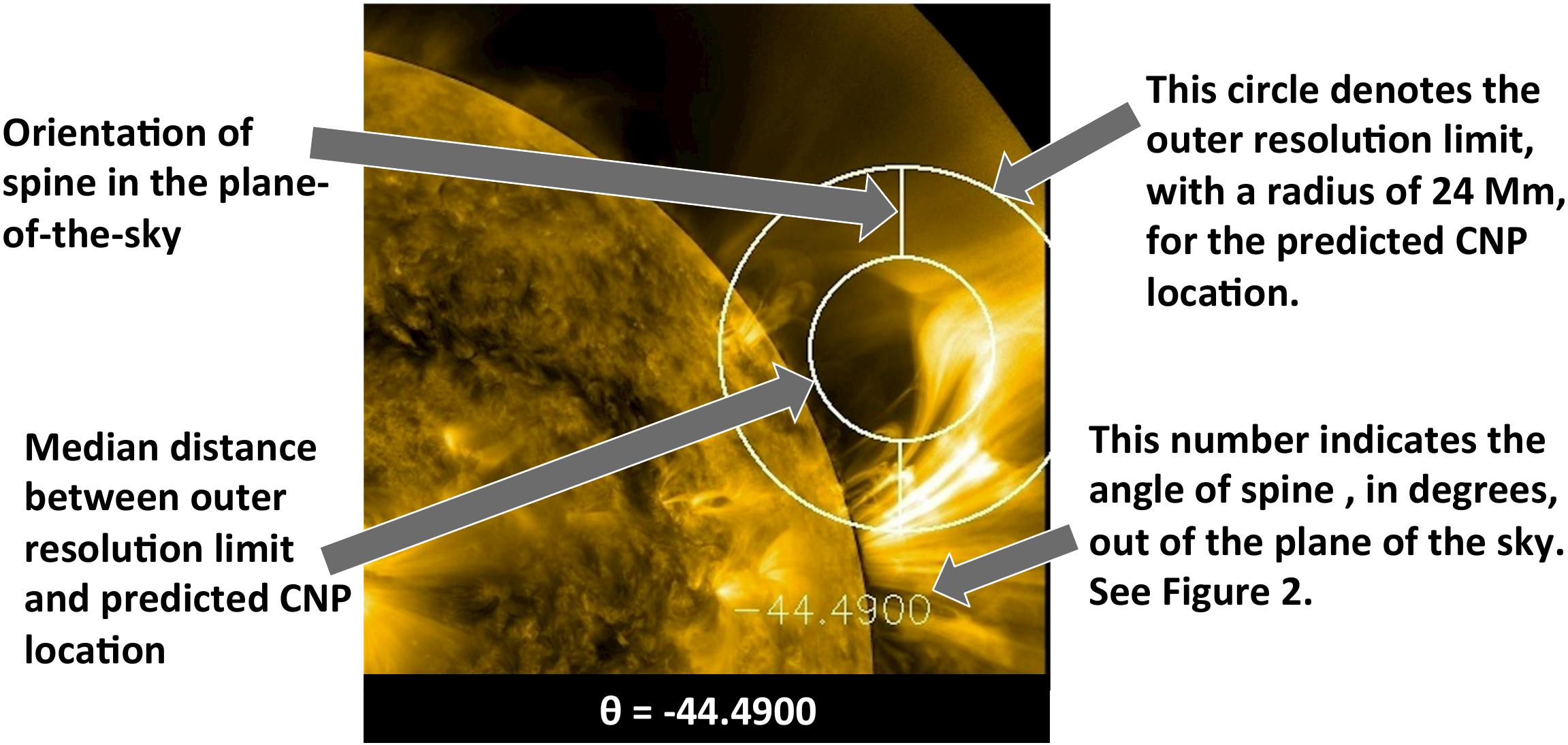}}
 \caption{An example of an error template overlaid onto a contrast-enhanced image produced with \changes{{\sf aia\_rfilter}}. The annulus shown above is centered at the predicted coronal null-point (CNP) location. An inner circle, with a radius half that of the outer circle, was included to help the viewer estimate how far off the observed structure might be from the predicted location. The theta value at the bottom is a clearer view of the spine angle given in the figure below the annulus.}
 \label{template_image}
 \end{figure}

\section{Methodology}
     \label{Methodology}
In this section we describe how the PFSS model was used to establish the locations of CNPs in the SDO/AIA field of view. Then, we detail how this information is used to download SDO/AIA images in the 171\,\AA\ band in order to conduct a manual inspection for possible CNP structures. This procedure includes applying a radial gradient (RG) filter to enhance the contrast in all of the 171\,\AA\ images to facilitate the visual inspection process. Also, an explanation of the additional information included on every contrast-enhanced image (Figure \ref{template_image}) is given. 
 
     \subsection{PFSS Model}
		\label{PFSS Model}
The PFSS model assumes a current-free magnetic field, which can be written as the gradient of a scalar potential, $[\textbf {\textit{B}}=-\nabla\chi]$.  The key assumption of the PFSS is that the field becomes purely radial at a radius called the {\em source surface} $[r=R_{\rm SS}]$.  This is achieved by setting $\chi=0$ there.  The harmonic scalar potential satisfying this is
\begin{equation}
  \chi ~=~ \sum_{\ell, m} P^{(m)}_{\ell}(\cos\theta)\left[\left({R_{\rm{SS}}\over r}\right)^{\ell+1}
  \!\!\!- \left({r\over R_{\rm{SS}}}\right)^{\ell}\right]\, \Bigl[ \,g^m_{\ell}\cos(m\phi) +h^m_{\ell}\sin(m\phi)\, \Bigr] ~~,
  	\label{eq:PFSS}
\end{equation}
where $P^{(m)}_{\ell}(x)$ is the associated Legendre function.  The real 
coefficients $g^m_{\ell}$ and $h^m_{\ell}$ are found by matching a synoptic magnetogram of a single Carrrington rotation built up from one month of line-of-sight magnetograms.  The potential field $\chi(r,\theta,\phi)$ then generates a static potential field, which we take to represent the field over the entire Carrington rotation.  We obtained, from the {\it Wilcox Solar Observatory} (WSO), one set of harmonic coefficients up to $\ell=29$ for each Carrington Rotation after rotation number 1893. (The WSO harmonic coefficients were obtained from \href{http://wso.stanford.edu/forms/prgs.html}{http://wso.stanford.edu/forms/prgs.html} courtesy of J.T. Hoeksema.)\par

For each Carrington rotation we performed a preliminary search for null points using a rectilinear grid covering radii $1.05\,\rm{R}_{\odot}\le r\le2.5\,\rm{R}_{\odot}$, with spacing of $3\,^{\circ}$ in latitude and longitude.  Expansion (\ref{eq:PFSS}), with  harmonics for each rotation, is used to compute the field vector ${\textbf {\textit{B}}}=(B_r,B_{\theta},B_{\phi})$ on each grid point.  Each of the rectilinear cells is then broken into five simplices (tetrahedra).    The field inside a single simplex is uniquely defined by linear interpolation from the values ${\textbf {\textit{B}}}_1,\, {\textbf {\textit{B}}}_2,\, {\textbf {\textit{B}}}_3$, and ${\textbf {\textit{B}}}_4$ at its four vertices. According to the algorithm of \citet{Greene92}, the interpolated field vanishes within the geometric simplex if the simplex defined by the four vectors ${\textbf {\textit{B}}}_1,\, {\textbf {\textit{B}}}_2,\, {\textbf {\textit{B}}}_3$, and ${\textbf {\textit{B}}}_4$, contains the origin $[{\textbf {\textit{B}}}=0]$.  For each simplex where this occurs we use the location of the interpolated null point to initialize a search, using Newton's method, for a genuine zero in the field defined by Equation\ (\ref{eq:PFSS}).  It is the null point found by this iterative scheme that we identify as one predicted null point in the PFSS for that Carrington Rotation. We have taken the field 
${\textbf {\textit{B}}}$ to represent an entire \changes{Carrington} rotation.  Any null point that we find will therefore cross the east and west limbs at precisely known times, which we henceforth identify with that null point.\par

Once the location of the null point ${[\textbf {\textit{x}}}_0]$, is found the expansion in Equation (\ref{eq:PFSS}) can be differentiated to compute the Jacobian matrix for that null point
\begin{equation}
  M_{ij}~=~ \left.{\partial B_i\over\partial x_j}\right\vert_{{\bf x}_0} ~=~ -
  \left.{\partial^2\chi\over\partial x_i\partial x_j}\right\vert_{{\bf x}_0}~~,
\end{equation}
where $x_i$ are Cartesian coordinates.  Since the matrix is symmetric and traceless \citep{Parnell96} it has three real eigenvalues, $[\lambda_1$, $\lambda_2$, and $\lambda_3]$, which must sum to zero.  One eigenvalue, designated $\lambda_1$, will have a sign opposite to the other two.  The corresponding eigenvector, ${[\bf \hat{\textbf {\textit{e}}}}_1]$, is the {\em spine} of the null point -- we resolve the directional ambiguity by defining ${\bf\hat{\textbf {\textit{r}}}}\cdot{\bf\hat{\textbf {\textit{e}}}}_1>0$ \changes{(where $\hat\textbf {\textit{r}}$ is a radial unit vector normal to the surface)}.  If $\lambda_1$ is positive (negative) the null point itself is designated negative (positive).  Thus for every null point we have a location, the eigenvalue $[\lambda_1]$, the spine direction $[{\hat{\textbf {\textit{e}}}}_1]$, and
ratio of the remaining eigenvalues, $[\lambda_3/\lambda_2]$, which we take to be in the range $(0,1)$.\par
\changes{This work focuses on the characteristics of the spine associated with the predicted CNP. However, future research will explore the equally important separatrix surfaces of the observed null points found in this article.}

\subsection{SDO/AIA Observations}
SDO/AIA is well suited for performing this type of survey due to its continuous coverage of the entire solar limb, high temporal cadence of 12 seconds, and angular resolution of 0.6 arcseconds $\rm pixel^{-1}$ \citep{Lemen12}. All observational work reported here focused on the 171\,\AA\ band, because of the greater contrast that it provides for resolving magnetic structures when compared to the other five EUV bands centered on iron lines. The sharper contrast of 171\,\AA\ could be the result of its distinct single peak and narrow temperature-response function as shown in Figure 11 of \citet{Lemen12}. Images were obtained via {\sf IDL SolarSoftWare} \citep{Freeland98} command {\sf vso\_get}.\par
Above-the-limb portions of the 171\,\AA\ images were enhanced using a radial gradient (RG) filter. The IDL program {\sf aia\_rfilter} can resolve faint features in the corona by combining multiple images of the region surrounding the solar disk which improves the signal-to-noise level. The corona is broken up into multiple concentric rings, which are scaled as a function of radius to increase the intensity further from the solar disk \citep{Masson14}. The photon flux is not conserved with this technique, but this will not be necessary for the work presented here. We are only interested in the morphology of structures and do not intend to perform any photometry.(See \href{http://aia.cfa.harvard.edu/rfilter.shtml}{http://aia.cfa.harvard.edu/rfilter.shtml} for the version of {\sf aia\_rfilter.pro} used here or to see examples of its effect on AIA images.)\par
 Five sequential images, starting at the predicted time that a null point crossed the east or west limb, were used as input for creating the filtered images. Occasionally there was trouble getting AIA data for a given time and instead of using RG filtering, the location was manually explored further with {\sf JHelioViewer} \citep{Mueller09} (The {\sf JHelioViewer} software can be obtained by visiting \href{http://jhelioviewer.org}{http://jhelioviewer.org}.)\par

\subsection{Additional Information on the Cataloged Images}
\begin{figure} [h]
 \centerline{\includegraphics[angle=-90,width=0.9\textwidth,clip=]{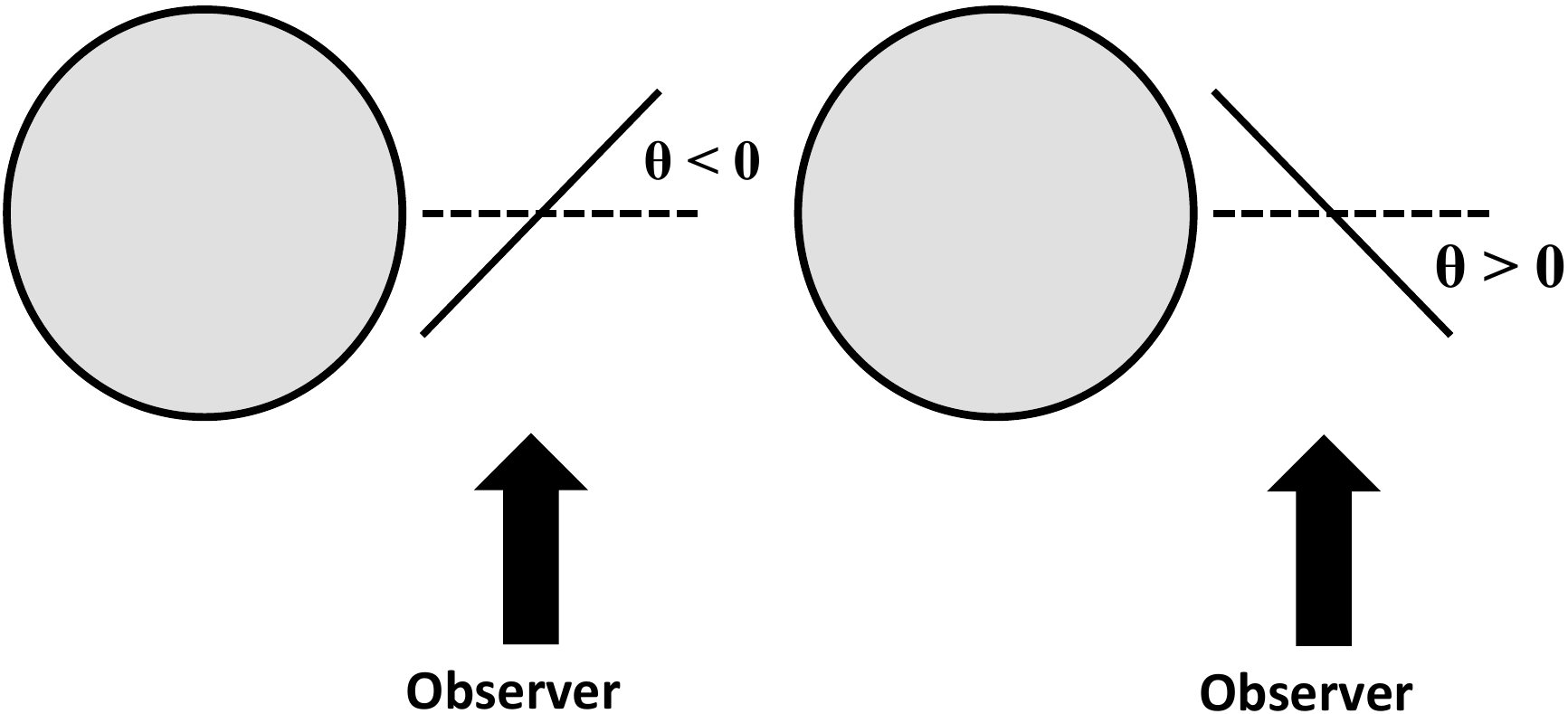}}
 \caption{This illustrates the meaning behind the numerical value included on the cataloged image shown in Figure \ref{template_image}. The number is associated with the spine's orientation angle, in degrees, out of the plane of the sky. Left and right sides show a configuration with a negative and positive $\theta$ value respectively.}
 \label{tilt_angle}
 \end{figure}

An error template, as shown in Figure \ref{template_image}, was overlaid on each enhanced image to indicate the predicted location of a CNP. The outer circle indicates the uncertainty from using harmonic coefficients up to only $\ell=29$ in Equation\ (\ref{eq:PFSS}) which translates to a radius $\approx\,24\,{\rm Mm}$. An inner circle, with a radius half that of the outer circle, was included to help the viewer estimate how far off the observed structure might be from the predicted location. Inside the annulus is a line indicating the direction of the spine projected onto the plane of the sky. The number shown on the enhanced images gives the angle in degrees, of the spine out of the plane of the sky. A negative value indicates that the line furthest from the solar disk is inclined away from the observer, while the line closer to the disk is inclined toward them. The opposite configuration would be true for a positive value, as shown in Figure \ref{tilt_angle}. \par
\begin{figure}[h]
 \centerline{\includegraphics[angle=-90,width=0.9\textwidth,clip=]{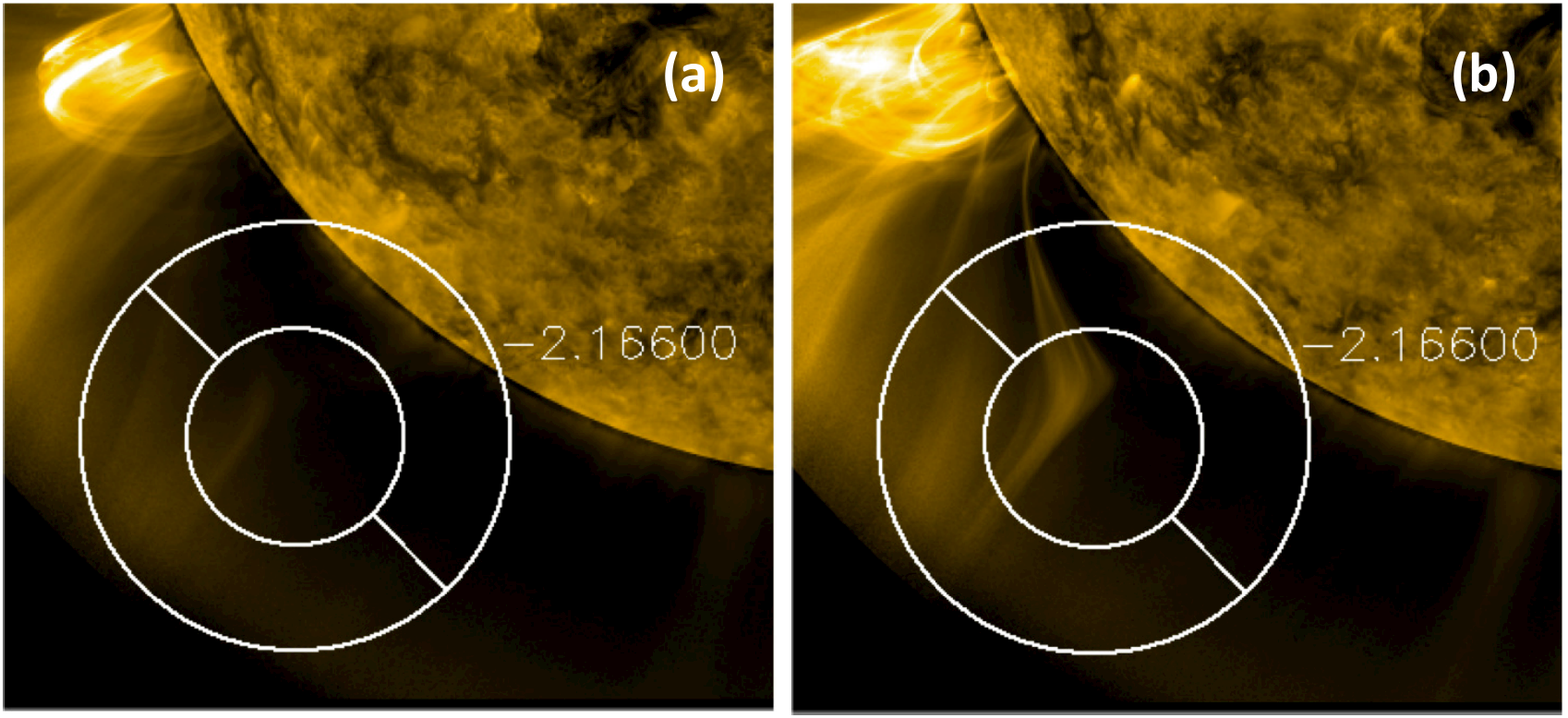}}
 \caption{An example of a CNP that was classified as not observed at the predicted time on 2012-05-06T23:15 (a), \changes{found to have an identifiable coronal structure} when investigated further with {\sf JHelioViewer} at 2012-05-07T13:30 (b).}
 \label{temporal_uncertainty}
 \end{figure}

An additional check was made with the use of {\sf JHelioViewer} to ensure that nulls were not missed due to any uncertainty in the predicted crossing time, as shown for example in Figure \ref{temporal_uncertainty}. Every CNP crossing was manually checked by examining images $\pm36$ hours from the predicted crossing time at a 25-minute cadence. Contrast enhancements were made by adjusting the gamma correction in {\sf JHelioViewer} when necessary. (Gamma is an exponent that is applied to the image array to make final results easier to see.) If this additional check finds a null at a slightly offset time, then a RG filter was applied to this new time and added to the catalog with the original prediction laid beside it for comparison as shown in Figure \ref{temporal_uncertainty}.\par

	Finally, a decision was made based on visual inspection as to whether or not a CNP structure was present at the predicted location. One of the requirements for a positive identification consisted of observing the structure inside the larger circle shown on the error template. \changes{For this work, we are only interested in identifying structures resembling a spine or separatrix surfaces of CNPs from SDO/AIA observations.} All of the {\sf jpeg} images that show the potential null-point locations, with an error template, can be found in our publicly available, on-line catalog of coronal null points at \href{http://solar.physics.montana.edu/mfreed/Null\_point\_research/data/}{http://solar.physics.montana.edu/mfreed/Null\_point\_research/data/} \changes{or with the on-line version of this journal article.} The data are separated into observed and non-observed CNPs for each limb.
	
\section{Results}
     \label{Results}
     The PFSS model predicted a total of 409 CNP locations, but nine were removed from the list for Carrington Rotations 2098 and 2099 due to limited availability of SDO/AIA data in the Summer of 2010. \changes{The region used for this study was determined by the PFSS model resolution (lower limit = $1.05\,\rm{R}_{\odot}$) and by SDO/AIA's field of view (upper limit = $1.30\,\rm{R}_{\odot}$)}. Only 294 of these locations were found within $1.05\,\rm{R}_{\odot} < r \leq 1.30\,\rm{R}_{\odot} $ which means that there were 588 chances to spot a CNP on either the east or west limb. Another six possible locations had to be removed due to unavailable data or off center images from SDO/AIA (i.e. due to AIA calibrations, engineering studies, or off pointing). This resulted in a total sample size of 582 for this catalog.\par
      {\sf JHelioViewer} had to be used for imaging 11 of the potential null-point locations because SDO/AIA FITS files were not available. These images do not have error templates like the ones obtained by applying the RG filter. Three other locations used RG filtering, but their limb crossing times were offset by a few hours due to missing AIA data.\par
\begin{figure} [H]
 \centerline{\includegraphics[width=0.8\textwidth,clip=]{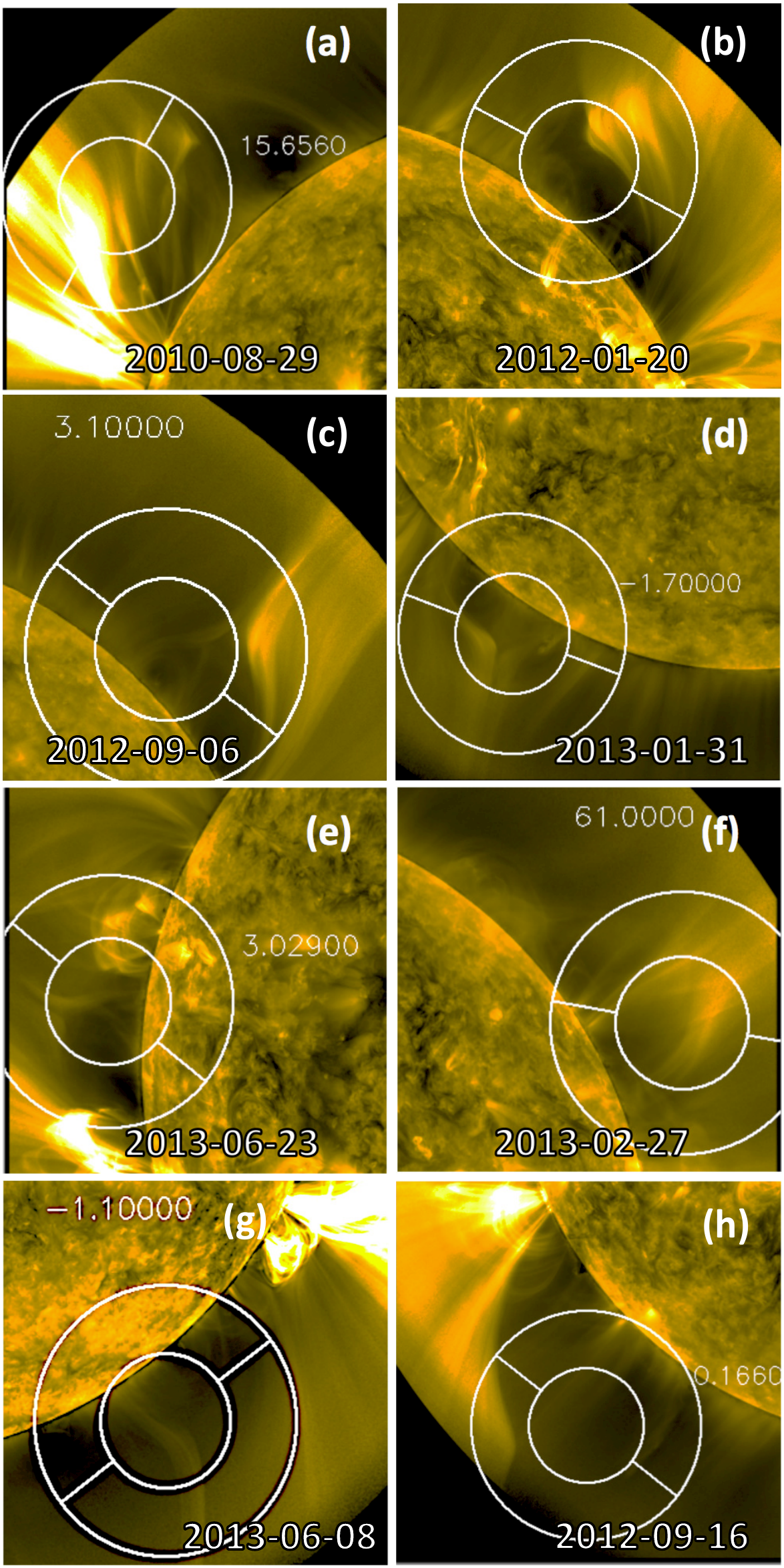}}
 \caption{Examples of observed CNPs from the compiled catalog. (a) and (b) are examples of asymmetric coronal streamers and (c)\,--\,(f) show symmetric coronal streamers. (b) also happens to be a CNP found and explored further in \protect\citet{Masson14}. (c) and (d) show the predicted spine orientation to be nearly perpendicular to the observed spine at the top of the dome and (e) illustrates an example of the two agreeing with each other quite well. (g) and (h) are examples of features described as ``mono-streamers" \changes{(description of this feature can be found in section 4.1)} that were included as a positive result. }
 \label{positive_find}
 \end{figure}

 
    Figure \ref{positive_find} shows some of the different kinds of features observed at the potential CNP sites. The most commonly observed structure at these locations were asymmetrical (Figure \ref{positive_find} a and b) and symmetrical coronal streamers (Figure \ref{positive_find} c\,--\,f). There were 33 images that also showed evidence of a spine (Figure \ref{positive_find} c\,--\,e). The spine either appears as a solid column of plasma above a cusp shaped loop top as shown by \citet{Sun12}, or as a clear channel of non-emitting plasma as shown in Figure \ref{positive_find} d . There is a $90\,^{\circ}$ discrepancy between the predicted and observed spine orientation in Figures \ref{positive_find} c and d. This could be due to the model predicting the incorrect spine orientation. \changes{However, since the predicted spine agrees more with the dome structure orientation, it is more conceivable that the spine is associated with the dome and the spike feature is related to the fan that is forming a separatrix curtain. Many of the observed CNPs appear very close to each other both spatially and temporally. The spines could be connecting to one of the neighboring CNP or bald patchs to form a null-null line as depicted in Figure 3 of \citet{Titov11}. Future work will investigate the structure of the separatrix surfaces to confirm this assertion, but it is beyond the scope of this work.}\par 
    
	The east limb had a total of 93 observed CNPs, and 31.8\,\% of CNPs predicted were observed; the west limb had 90 observed CNPs, and 30.3\,\% predicted were observed. There were 48 predicted CNPs observed on both limbs. All 48 of these filtered images were combined side-by-side for easier comparison, and can be found at the before mentioned website in Section 2.3 \changes{or with the on-line version of this journal article.} A total of 39 CNPs were reclassified as observed after a second examination with {\sf JHelioViewer} as mentioned in Section 2.2. This means that \changes{93.3\,\% of the predicted CNPs, or 543 out of the 582 in the catalog, were correctly identified before an additional check was conducted with JHelioViewer.}\par 
		Information regarding all CNPs observed on either solar limb can be found in Table \ref{tbl:1}. This table includes east and west limb crossing time, which limb shows evidence of structure, latitude of CNP at limb crossing, radial distance, sign of null point, ratio of the eigenvalues that form the fan, eigenvalue associated with the spine, and tilt of spine in and out of the plane of the sky. Note that the spine's tilt angle out of the plane of the sky will have the opposite sign, as given in Table \ref{tbl:1}, when projected onto the east limb. A complete list, which includes CNPs not observed on either limb, can be found in our on-line CNP catalog.
		
\begin{landscape}
\begin{center}
\tiny
\begin{longtable}{>{\raggedleft\arraybackslash} m{0.78cm} >{\raggedleft\arraybackslash} m{2.3cm} >{\raggedleft\arraybackslash} m{2.3cm} >{\raggedleft\arraybackslash} m{1.1cm} >{\raggedleft\arraybackslash} m{0.95cm} >{\raggedleft\arraybackslash} m{0.77cm} >{\raggedleft\arraybackslash} m{1.22cm} >{\raggedleft\arraybackslash} m{0.6cm} >{\raggedleft\arraybackslash} m{1.0cm} >{\raggedleft\arraybackslash} m{0.8cm} >{\raggedleft\arraybackslash} m{1.4cm} >{\raggedleft\arraybackslash} m{1.4cm}} 

\caption{Coronal null points that appear to have structure as they crossed the east, west or both limbs of the Sun. Complete list of predicted null points can be found at \href{www.solar.physics.montana.edu/mfreed/Null\_point\_research/data/}{www.solar.physics.montana.edu/mfreed/Null\_point\_research/data/} }
\label{tbl:1}\\
\hline
Line No. & East Limb Crossing & West Limb Crossing & Structure on Limb & Latitude & Radius & Carrington Rotation & Sign of Null Point & \changes{Ratio of Same Sign Eigenvalue} & Odd Eigenvalue & Tilt of Spine in the Plane with Respect to Vector Normal to Surface & Tilt of Spine In/Out of the Plane of the Sky on the West Limb \\
\hline
\endfirsthead
 
\caption{Continues}\\
\hline 
Line No. & East Limb Crossing & West Limb Crossing & Structure on Limb & Latitude & Radius & Carrington Rotation & Sign of Null Point & \changes{Ratio of Same Sign Eigenvalue} & Odd Eigenvalue & Tilt of Spine in the Plane with Respect to Vector Normal to Surface & Tilt of Spine In/Out of the Plane of the Sky on the West Limb\\
\hline
\endhead
\multicolumn{12}{c}
\textbf{Continues on the next page $\rightarrow$}\\
\endfoot
\hline
\endlastfoot

1 & 2010-06-25T10:46 & 2010-07-08T22:35 & West & -13.6 & 1.056 & 2098 & (-) & 0.19 & 7.25 & 27.42 & -49.48\\
2 & 2010-07-16T14:54 & 2010-07-29T12:16 & East & -42.8 & 1.073 & 2099 & (+) & 0.20 & -2.89 & -76.30 & -6.41 \\
3 & 2010-07-25T12:05 & 2010-08-08T19:58 & Both & 39.3 & 1.096 & 2099 & (-) & 0.72 & 5.61 & 71.90 & -10.05 \\
4 & 2010-08-24T05:48 & 2010-09-04T06:00 & Both & -66.7 & 1.073 & 2100 & (-) & 0.38 & 3.49 & 50.02 & 3.66 \\
5 & 2010-08-28T11:26 & 2010-09-11T20:03 & East & 33.4 & 1.190 & 2100 & (+) & 0.48 & -1.42 & -13.75 & -73.04\\
6 & 2010-08-29T10:58 & 2010-09-12T22:13 & Both & 37.2 & 1.286 & 2100 & (-) & 0.38 & 0.93 & 81.83 & -15.66 \\
7 & 2010-08-29T20:15 & 2010-09-13T14:15 & Both & 45.2 & 1.141 & 2101 & (-) & 0.57 & 2.43 & 88.51 & -1.60 \\
8 & 2010-09-04T22:24 & 2010-09-19T22:41 & West & 51.3 & 1.059 & 2101 & (-) & 0.37 & 7.37 & 58.01 & -13.36\\
9 & 2010-09-10T15:33 & 2010-09-23T02:09 & Both & -47.6 & 1.060 & 2101 & (-) & 0.02 & 6.28 & -10.35 & -0.46\\
10 & 2010-09-14T21:33 & 2010-09-27T13:41 & West & -41.7 & 1.077 & 2101 & (+) & 0.37 & -5.90 & -74.79 & -33.42 \\
11 & 2010-09-26T08:26 & 2010-10-10T22:24 & East & 43.1 & 1.138 & 2102 & (-) & 0.81 & 2.60 & 82.42 & 1.20 \\
12 & 2010-10-08T04:07 & 2010-10-20T16:24 & Both & -51.3 & 1.074 & 2102 & (-) & 0.19 & 4.05 & -8.01 & 32.58 \\
13 & 2010-10-12T08:26 & 2010-10-26T18:45 & West & 42.6 & 1.245 & 2102 & (-) & 0.30 & 0.67 & 67.62 & 27.21 \\
14 & 2010-10-12T19:41 & 2010-10-27T04:41 & East & 40.8 & 1.080 & 2102 & (+) & 0.83 & -3.39 & -74.00 & 17.58 \\
15 & 2010-10-18T09:45 & 2010-10-31T22:35 & East & -8.0 & 1.083 & 2102 & (-) & 0.48 & 3.45 & 47.22 & -43.45 \\
16 & 2010-11-14T09:56 & 2010-11-27T22:58 & East & -18.2 & 1.248 & 2103 & (-) & 0.42 & 0.53 & 68.66 & -18.40 \\
17 & 2010-12-26T19:30 & 2011-01-09T20:15 & West & -36.5 & 1.052 & 2105 & (-) & 0.66 & 4.62 & 16.31 & -42.37 \\
18 & 2011-01-07T01:24 & 2011-01-19T21:33 & East & 51.5 & 1.135 & 2105 & (-) & 0.39 & 1.57 & -82.28 & 14.12 \\
19 & 2011-01-08T15:00 & 2011-01-22T08:15 & West & -4.1 & 1.070 & 2105 & (+) & 0.26 & -1.86 & -77.46 & 1.18 \\
20 & 2011-02-13T01:24 & 2011-02-27T05:48 & West & -26.0 & 1.085 & 2107 & (+) & 0.38 & -1.11 & -21.13 & 21.77 \\
21 & 2011-02-19T16:01 & 2011-03-06T07:46 & Both & -42.3 & 1.074 & 2107 & (+) & 0.13 & -2.62 & -68.07 & 5.66 \\
22 & 2011-04-01T14:48 & 2011-04-15T05:37 & West & 1.4 & 1.075 & 2108 & (+) & 0.09 & -4.80 & -86.80 & -7.89 \\
23 & 2011-04-03T01:18 & 2011-04-16T15:56 & West & 1.8 & 1.089 & 2108 & (+) & 0.48 & -3.62 & -6.32 & -81.59 \\
24 & 2011-04-20T19:24 & 2011-05-04T06:22 & Both & 13.6 & 1.095 & 2109 & (+) & 0.51 & -5.88 & -0.32 & -86.73 \\
25 & 2011-04-30T17:09 & 2011-05-14T08:20 & Both & -2.4 & 1.074 & 2109 & (+) & 0.26 & -3.11 & -5.45 & -66.05 \\
26 & 2011-05-06T22:35 & 2011-05-20T15:11 & West & -10.3 & 1.143 & 2110 & (+) & 0.56 & -1.34 & 62.89 & -43.23 \\
27 & 2011-05-16T04:13 & 2011-05-29T18:16 & Both & 3.3 & 1.057 & 2110 & (+) & 0.26 & -7.37 & -31.09 & -57.55 \\
28 & 2011-05-20T08:09 & 2011-06-03T00:22 & Both & -20.4 & 1.106 & 2110 & (-) & 0.26 & 4.36 & 25.55 & -68.28 \\
29 & 2011-05-21T08:09 & 2011-06-03T23:37 & Both & -13.1 & 1.176 & 2110 & (+) & 0.59 & -2.93 & -8.55 & 57.97 \\
30 & 2011-06-01T12:39 & 2011-06-15T02:48 & East & -32.2 & 1.093 & 2111 & (+) & 0.54 & -2.66 & -69.67 & -1.29 \\
31 & 2011-06-06T07:13 & 2011-06-19T20:54 & East & -13.6 & 1.142 & 2111 & (+) & 0.54 & -2.40 & 38.91 & 63.52 \\
32 & 2011-06-23T12:11 & 2011-07-07T06:39 & Both & 22.8 & 1.103 & 2111 & (+) & 0.16 & -9.19 & -8.08 & 43.18 \\
33 & 2011-07-02T17:48 & 2011-07-16T12:39 & East & 17.7 & 1.052 & 2112 & (-) & 0.35 & 2.14 & 36.46 & 54.97 \\
34 & 2011-07-08T06:28 & 2011-07-22T15:11 & West & 49.3 & 1.225 & 2112 & (-) & 0.52 & 1.16 & -87.86 & 10.40 \\
35 & 2011-07-22T04:18 & 2011-08-05T01:24 & East & 18.0 & 1.064 & 2112 & (-) & 0.13 & 20.47 & -2.95 & 87.95 \\
36 & 2011-09-09T23:43 & 2011-09-25T00:00 & East & 51.5 & 1.144 & 2114 & (-) & 0.87 & 2.41 & 89.95 & -0.28 \\
37 & 2011-09-11T18:56 & 2011-09-25T08:09 & East & -4.5 & 1.102 & 2114 & (-) & 0.62 & 6.30 & 54.69 & -33.34 \\
38 & 2011-09-17T04:10 & 2011-10-02T05:03 & West & 52.7 & 1.094 & 2115 & (-) & 0.25 & 3.36 & 84.02 & 7.50 \\
39 & 2011-09-30T01:24 & 2011-10-14T15:28 & East & 43.8 & 1.082 & 2115 & (-) & 0.43 & 3.91 & -85.82 & -2.60 \\
40 & 2011-10-07T08:26 & 2011-10-22T06:16 & West & 54.3 & 1.159 & 2115 & (-) & 0.74 & 2.34 & -80.48 & 8.32 \\
41 & 2011-10-10T17:09 & 2011-10-23T14:03 & East & -41.5 & 1.244 & 2115 & (+) & 0.89 & -1.53 & -65.51 & -03.88 \\
42 & 2011-10-15T04:10 & 2011-10-29T21:16 & East & 52.8 & 1.073 & 2116 & (-) & 0.08 & 4.16 & 76.13 & 1.47 \\
43 & 2011-10-21T18:28 & 2011-11-04T18:39 & Both & 26.6 & 1.162 & 2116 & (+) & 0.42 & -5.61 & 63.13 & -44.49 \\
44 & 2011-10-22T10:07 & 2011-11-05T12:33 & Both & 32.3 & 1.094 & 2116 & (-) & 0.51 & 6.66 & 27.31 & 40.60 \\
45 & 2011-10-22T21:56 & 2011-11-05T06:33 & West & -22.3 & 1.157 & 2116 & (+) & 0.39 & -1.37 & -79.07 & -26.26 \\
46 & 2011-10-23T10:52 & 2011-11-05T18:28 & West & -25.1 & 1.185 & 2116 & (-) & 0.52 & 1.22 & 15.40 & -13.44 \\
47 & 2011-10-23T18:22 & 2011-11-06T22:30 & West & 37.1 & 1.060 & 2116 & (+) & 0.62 & -8.73 & -20.59 & -76.05 \\
48 & 2011-11-07T16:01 & 2011-11-20T21:28 & East & -42.9 & 1.194 & 2116 & (+) & 0.79 & -1.94 & -69.49 & 1.27 \\
49 & 2011-12-13T09:45 & 2011-12-27T06:39 & West & -42.3 & 1.087 & 2118 & (+) & 0.28 & -5.30 & -65.69 & 15.29 \\
50 & 2011-12-14T08:09 & 2011-12-28T01:13 & West & -9.7 & 1.106 & 2118 & (-) & 0.49 & 5.35 & -56.92 & 42.13 \\
51 & 2011-12-15T10:13 & 2011-12-28T23:54 & West & 19.1 & 1.194 & 2118 & (+) & 0.11 & -3.25 & 72.09 & -6.39 \\
52 & 2011-12-15T12:05 & 2011-12-29T01:30 & West & 20.1 & 1.134 & 2118 & (-) & 0.10 & 7.17 & -2.04 & 88.47 \\
53 & 2011-12-27T02:15 & 2012-01-09T19:07 & East & -4.2 & 1.162 & 2118 & (-) & 0.24 & 5.32 & 41.56 & 55.10 \\
54 & 2012-01-07T05:37 & 2012-01-20T01:01 & Both & 52.3 & 1.111 & 2119 & (-) & 0.45 & 3.40 & -80.04 & 2.85 \\
55 & 2012-01-08T04:24 & 2012-01-21T00:45 & Both & 50.5 & 1.064 & 2119 & (+) & 0.22 & -5.50 & 8.93 & 70.65 \\
56 & 2012-01-12T11:20 & 2012-01-26T10:58 & West & -22.2 & 1.259 & 2119 & (+) & 0.13 & -1.48 & 17.41 & 4.32 \\
57 & 2012-01-29T12:00 & 2012-02-11T16:24 & West & 27.2 & 1.127 & 2119 & (+) & 0.81 & -6.17 & -41.06 & 61.52 \\
58 & 2012-02-04T12:50 & 2012-02-18T03:39 & Both & 3.1 & 1.079 & 2120 & (-) & 0.65 & 3.69 & 15.24 & -76.58 \\
59 & 2012-02-09T01:18 & 2012-02-23T07:46 & Both & -29.8 & 1.057 & 2120 & (+) & 0.27 & -6.63 & -12.77 & 54.01 \\
60 & 2012-02-13T07:18 & 2012-02-26T15:50 & East & 16.4 & 1.133 & 2120 & (+) & 0.09 & -2.37 & 45.88 & 51.63 \\
61 & 2012-02-16T12:56 & 2012-03-01T07:18 & East & -5.0 & 1.090 & 2120 & (+) & 0.29 & -6.89 & -50.05 & -52.49 \\
62 & 2012-02-19T08:43 & 2012-03-04T03:05 & West & -5.4 & 1.080 & 2120 & (-) & 0.31 & 8.49 & 56.15 & 47.48 \\
63 & 2012-02-22T10:10 & 2012-03-07T03:56 & West & -4.3 & 1.099 & 2120 & (+) & 0.20 & -5.28 & -17.64 & 2.92 \\
64 & 2012-03-01T16:52 & 2012-03-14T20:31 & West & 24.6 & 1.101 & 2121 & (-) & 0.19 & 14.37 & 1.20 & 88.94 \\
65 & 2012-03-05T20:54 & 2012-03-19T05:03 & East & 16.4 & 1.191 & 2121 & (-) & 0.75 & 2.43 & 6.58 & -85.24 \\
66 & 2012-03-28T23:37 & 2012-04-11T23:54 & West & -21.3 & 1.157 & 2122 & (+) & 0.10 & -3.04 & -35.44 & -56.52 \\
67 & 2012-04-09T11:20 & 2012-04-24T00:33 & Both & -48.2 & 1.211 & 2122 & (+) & 0.51 & -1.64 & -80.36 & 7.65 \\
68 & 2012-04-16T17:15 & 2012-04-30T08:48 & Both & -2.3 & 1.066 & 2122 & (+) & 0.85 & -7.55 & -26.48 & 27.19 \\
69 & 2012-04-18T22:24 & 2012-05-03T02:48 & Both & -38.2 & 1.115 & 2122 & (+) & 0.90 & -4.11 & -66.45 & -1.68 \\
70 & 2012-04-24T18:45 & 2012-05-08T00:28 & Both & 31.6 & 1.049 & 2123 & (+) & 0.16 & -7.60 & 81.83 & -23.75 \\
71 & 2012-04-27T17:31 & 2012-05-11T04:41 & East & 14.7 & 1.182 & 2123 & (-) & 0.16 & 1.98 & -10.38 & 81.52 \\
72 & 2012-05-02T04:10 & 2012-05-16T01:13 & East & -28.1 & 1.218 & 2123 & (+) & 0.52 & -2.05 & -20.26 & 42.99 \\
73 & 2012-05-03T08:54 & 2012-05-16T19:52 & Both & 17.5 & 1.090 & 2123 & (-) & 0.12 & 8.12 & 39.10 & -59.40 \\
74 & 2012-05-06T23:15 & 2012-05-21T02:03 & Both & -50.6 & 1.220 & 2123 & (+) & 0.65 & -2.13 & -84.21 & 2.17 \\
75 & 2012-05-11T08:03 & 2012-05-24T22:13 & Both & 4.1 & 1.068 & 2123 & (-) & 0.30 & 13.30 & -8.50 & 83.31 \\
76 & 2012-05-13T19:30 & 2012-05-27T09:22 & East & 5.9 & 1.210 & 2123 & (+) & 0.78 & -4.48 & -3.15 & 71.90 \\
77 & 2012-05-18T11:43 & 2012-06-01T02:48 & East & -7.3 & 1.270 & 2123 & (+) & 0.62 & -0.67 & -46.91 & 58.06 \\
78 & 2012-05-28T18:05 & 2012-06-11T09:00 & West & -31.7 & 1.141 & 2124 & (+) & 0.56 & -4.25 & -48.18 & 19.14 \\
79 & 2012-06-13T07:30 & 2012-06-26T21:50 & West & -1.6 & 1.094 & 2124 & (-) & 0.38 & 3.00 & -14.07 & -69.51 \\
80 & 2012-06-14T19:07 & 2012-06-28T03:39 & Both & -40.9 & 1.055 & 2124 & (-) & 0.38 & 5.25 & -0.36 & -89.04 \\
81 & 2012-06-19T06:00 & 2012-07-02T18:56 & West & -10.5 & 1.083 & 2125 & (+) & 0.46 & -7.85 & -11.35 & -81.31 \\
82 & 2012-06-19T11:48 & 2012-07-03T03:45 & West & 10.3 & 1.213 & 2125 & (-) & 0.35 & 1.28 & -2.89 & 85.22 \\
83 & 2012-06-20T20:48 & 2012-07-04T05:54 & East & -30.0 & 1.122 & 2125 & (-) & 0.58 & 6.61 & 38.98 & 60.59 \\
84 & 2012-06-26T22:46 & 2012-07-10T12:11 & East & -5.6 & 1.188 & 2125 & (-) & 0.51 & 6.62 & 71.50 & -27.47 \\
85 & 2012-06-26T23:54 & 2012-07-11T15:56 & East & 65.2 & 1.183 & 2125 & (-) & 0.50 & 0.75 & -72.04 & -8.19 \\
86 & 2012-06-28T00:39 & 2012-07-12T12:22 & East & 60.4 & 1.090 & 2125 & (+) & 0.26 & -2.15 & -11.95 & 66.15 \\
87 & 2012-07-06T01:35 & 2012-07-19T23:43 & East & 27.1 & 1.073 & 2125 & (+) & 0.62 & -9.78 & 72.21 & -8.28 \\
88 & 2012-07-19T08:43 & 2012-08-01T11:20 & East & -31.5 & 1.132 & 2126 & (-) & 0.23 & 4.65 & 43.37 & 52.94 \\
89 & 2012-08-01T14:26 & 2012-08-12T23:15 & Both & -66.9 & 1.125 & 2126 & (-) & 0.69 & 3.20 & 24.71 & 6.49 \\
90 & 2012-08-19T04:01 & 2012-08-31T15:11 & Both & -46.8 & 1.221 & 2127 & (+) & 0.16 & -1.76 & -84.80 & -2.36 \\
91 & 2012-08-22T11:54 & 2012-09-06T05:15 & Both & 44.9 & 1.080 & 2127 & (+) & 0.75 & -5.80 & -83.54 & 3.07 \\
92 & 2012-08-29T22:18 & 2012-09-12T15:50 & West & 5.2 & 1.090 & 2127 & (-) & 0.64 & 5.92 & -1.06 & 88.85 \\
93 & 2012-08-30T16:41 & 2012-09-10T13:46 & Both & -67.4 & 1.182 & 2127 & (-) & 0.14 & 3.25 & 20.54 & 13.06 \\
94 & 2012-08-31T01:58 & 2012-09-11T12:22 & Both & -63.1 & 1.112 & 2127 & (+) & 0.10 & -5.46 & 73.68 & -28.89 \\
95 & 2012-09-02T04:18 & 2012-09-15T08:54 & Both & -21.6 & 1.053 & 2127 & (-) & 0.79 & 16.99 & -51.86 & -42.88 \\
96 & 2012-09-02T20:54 & 2012-09-15T23:26 & Both & -25.7 & 1.054 & 2127 & (+) & 0.33 & -14.35 & 5.08 & 78.17 \\
97 & 2012-09-16T01:58 & 2012-09-28T11:15 & East & -49.1 & 1.260 & 2128 & (+) & 0.43 & -1.26 & 87.54 & -0.17 \\
98 & 2012-09-20T22:01 & 2012-10-04T12:39 & West & -1.7 & 1.062 & 2128 & (-) & 0.59 & 9.44 & -54.78 & -38.07 \\
99 & 2012-09-27T19:41 & 2012-10-11T21:05 & West & 22.6 & 1.075 & 2128 & (+) & 0.63 & -2.86 & 42.15 & -25.09 \\
100 & 2012-10-09T04:52 & 2012-10-24T22:01 & Both & 67.2 & 1.307 & 2129 & (-) & 0.75 & 0.33 & -51.38 & -5.64 \\
101 & 2012-11-06T13:41 & 2012-11-20T04:01 & West & -6.7 & 1.066 & 2130 & (-) & 0.26 & 12.65 & 81.22 & 16.41 \\
102 & 2012-11-11T00:05 & 2012-11-23T22:30 & Both & -62.1 & 1.086 & 2130 & (-) & 0.45 & 3.57 & 43.15 & 40.56 \\
103 & 2012-11-17T16:07 & 2012-12-01T11:43 & West & 32.3 & 1.142 & 2130 & (-) & 0.16 & 1.38 & 4.72 & 80.55 \\
104 & 2012-12-03T22:30 & 2012-12-17T13:41 & East & 27.6 & 1.081 & 2131 & (+) & 0.34 & -9.09 & 29.31 & 11.04 \\
105 & 2012-12-04T08:54 & 2012-12-18T02:26 & East & -43.7 & 1.055 & 2131 & (+) & 0.58 & -4.98 & 53.32 & -51.62 \\
106 & 2012-12-11T06:28 & 2012-12-24T22:52 & West & -6.6 & 1.074 & 2131 & (-) & 0.66 & 5.57 & -76.39 & 14.85 \\
107 & 2012-12-12T03:50 & 2012-12-25T13:52 & Both & 48.7 & 1.197 & 2131 & (+) & 0.19 & -0.94 & 89.63 & -2.83 \\
108 & 2012-12-17T01:01 & 2012-12-30T20:20 & East & -24.6 & 1.213 & 2131 & (+) & 0.70 & -2.12 & 5.33 & 41.47 \\
109 & 2012-12-22T15:56 & 2013-01-05T14:31 & West & -33.9 & 1.113 & 2132 & (-) & 0.67 & 4.96 & -59.76 & -38.36 \\
110 & 2012-12-27T03:16 & 2013-01-09T17:31 & East & 8.3 & 1.154 & 2132 & (+) & 0.21 & -4.08 & 11.43 & -72.33 \\
111 & 2013-01-03T07:07 & 2013-01-17T19:52 & Both & -54.1 & 1.088 & 2132 & (+) & 0.19 & -4.53 & 88.27 & -2.53 \\
112 & 2013-01-19T05:31 & 2013-02-02T11:26 & East & -33.9 & 1.056 & 2133 & (-) & 0.63 & 7.70 & -58.88 & -26.97 \\
113 & 2013-01-21T00:05 & 2013-02-04T05:37 & East & -32.8 & 1.054 & 2133 & (+) & 0.37 & -5.72 & 65.56 & 6.34 \\
114 & 2013-01-28T14:31 & 2013-02-11T18:39 & Both & -27.8 & 1.130 & 2133 & (-) & 0.45 & 6.59 & 7.40 & -63.44 \\
115 & 2013-01-31T06:28 & 2013-02-15T12:00 & Both & -57.7 & 1.079 & 2133 & (+) & 0.14 & -4.65 & 77.94 & 1.70 \\
116 & 2013-02-01T06:45 & 2013-02-16T12:39 & Both & -57.6 & 1.086 & 2133 & (-) & 0.09 & 4.19 & -11.70 & 12.43 \\
117 & 2013-02-04T02:43 & 2013-02-18T05:03 & Both & -23.0 & 1.143 & 2133 & (+) & 0.41 & -3.65 & 67.75 & 35.81 \\
118 & 2013-02-14T13:07 & 2013-02-27T16:18 & Both & 26.3 & 1.183 & 2133 & (-) & 0.22 & 1.53 & -36.61 & 60.98 \\
119 & 2013-02-18T17:48 & 2013-03-05T08:09 & Both & -40.4 & 1.058 & 2134 & (+) & 0.15 & -5.28 & -6.64 & 21.72 \\
120 & 2013-02-21T21:33 & 2013-03-08T05:37 & West & -31.5 & 1.062 & 2134 & (+) & 0.23 & -6.99 & -8.61 & 78.38 \\
121 & 2013-03-07T06:28 & 2013-03-16T20:26 & East & 74.0 & 1.196 & 2134 & (-) & 0.97 & 1.59 & 5.74 & -41.15 \\
122 & 2013-03-18T14:20 & 2013-04-01T19:24 & Both & -28.3 & 1.075 & 2135 & (+) & 0.29 & -6.95 & 7.63 & -52.64 \\
123 & 2013-04-11T04:13 & 2013-04-25T04:13 & West & -24.2 & 1.116 & 2136 & (+) & 0.55 & -4.79 & -4.83 & 86.11 \\
124 & 2013-04-16T02:09 & 2013-04-29T21:22 & East & -12.5 & 1.171 & 2136 & (+) & 0.63 & -2.27 & 8.88 & -37.34 \\
125 & 2013-05-04T06:28 & 2013-05-16T21:11 & Both & 64.6 & 1.145 & 2136 & (-) & 0.58 & 1.52 & -28.13 & 20.64 \\
126 & 2013-05-17T20:31 & 2013-05-31T09:16 & East & 18.2 & 1.066 & 2137 & (+) & 0.09 & -8.25 & -6.63 & -57.47 \\
127 & 2013-05-19T02:43 & 2013-06-01T21:11 & West & -38.3 & 1.081 & 2137 & (-) & 0.83 & 6.65 & 68.50 & 26.60 \\
128 & 2013-05-25T14:43 & 2013-06-08T08:26 & West & -55.9 & 1.119 & 2137 & (-) & 0.40 & 2.68 & -87.49 & -1.06 \\
129 & 2013-06-02T07:18 & 2013-06-15T19:35 & East & -62.5 & 1.088 & 2137 & (+) & 0.13 & -1.62 & 60.31 & -23.28 \\
130 & 2013-06-06T05:54 & 2013-06-19T14:03 & East & -65.8 & 1.220 & 2138 & (+) & 0.32 & -1.02 & 41.40 & -13.08 \\
131 & 2013-06-23T08:48 & 2013-07-07T00:22 & East & 7.0 & 1.082 & 2138 & (+) & 0.34 & -6.00 & 31.83 & -3.03 \\
132 & 2013-06-26T14:26 & 2013-07-10T05:09 & Both & 1.3 & 1.205 & 2138 & (-) & 0.35 & 1.33 & -14.44 & -56.66 \\
133 & 2013-07-04T15:50 & 2013-07-17T19:18 & Both & -37.2 & 1.081 & 2139 & (+) & 0.36 & -4.35 & -27.06 & -12.52 \\
134 & 2013-07-04T22:41 & 2013-07-18T05:54 & West & -26.8 & 1.068 & 2139 & (-) & 0.21 & 5.61 & 36.87 & -63.44 \\
135 & 2013-07-23T22:41 & 2013-08-04T11:37 & East & -67.4 & 1.054 & 2139 & (+) & 0.26 & -5.42 & 28.72 & 15.29 \\
 
\hline
\end{longtable}
\end{center}
\end{landscape}

  \begin{figure} 
 \centerline{\includegraphics[angle=-90,width=0.9\textwidth,clip=]{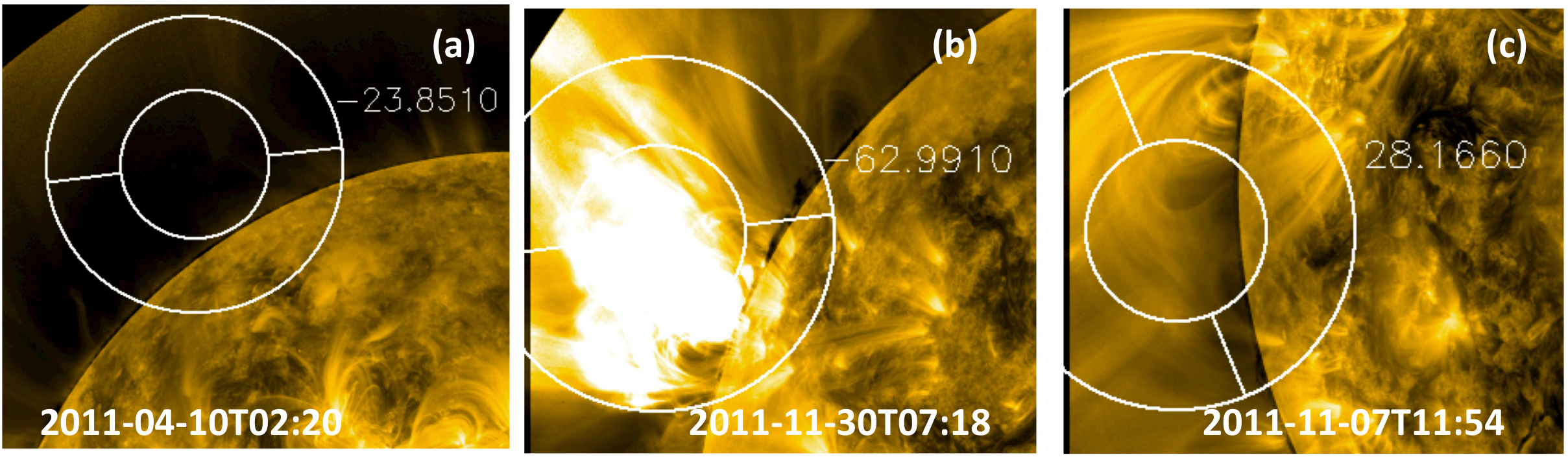}}
 \caption{Some examples of CNPs that were classified as not being observed at a predicted location as the result of (a) no emitting plasma, (b) image saturation, or (c) undistinguishable structure.}
 \label{negative_find}
 \end{figure}

\section{Discussion}
     \label{Discussion}
     \subsection{Catalog}
     Constituents of the CNP catalog consist of 35 symmetrical coronal streamers, 55 asymmetrical coronal streamers, and an additional 49.1\,\% that could not be clearly classified as either. Many of the asymmetric observations resembled structures found from numerical results in Figures 2 and 3 of \citet{Moreno13}. The symmetrical and asymmetrical coronal streamers resemble helmet and psuedo-streamers respectively. However,  \citet{Titov12} and \citet{Platten14} illustrates the topology investigation needed for making such a determination from observational data, which is beyond the scope of the work presented here. Therefore, the features are merely classified as symmetric or asymmetric when possible as shown in Figure \ref{positive_find}. Another feature is illustrated in Figures \ref{positive_find} g and h and is referred to as a ``mono-streamer". \changes{This nomenclature is used here for a configuration that appears to have only one leg of a dome structure emitting plasma. These features are usually seen at high latitudes and are most likely associated with the boundary of a polar corona hole.} The difficulty in discerning an observable CNP structure is emphasized in Figure \ref{positive_find}, where orientation (panel f) or low intensity (panel g) of CNP location can play a major role in classification.\par
     A  CNP structure was considered not present inside the error template if it resembled images shown in Figure \ref{negative_find}. This includes: (panel a) no emitting plasma present, (panel b) image saturation, or (panel c) undistinguishable structure at the predicted location. Regions absent of plasma emission tend to be located at high latitudes and comprised of 8.2\,\% of the predicted CNP population. \changes{The probability} of image saturation increased as a result of using RG filtering, which was the cause for eliminating 18.6\,\% of the \changes{total number of predicted CNPs}. This means that the reported percentages of observed structures should be taken as a lower limit due to the possibility of under-counting from saturation and non-emitting plasma locations.\par
      Our survey provides insight into the lifetimes of visible CNPs. A predicted CNP has a 16.7\,\% chance of being spotted on the east limb and later observed surviving its fourteen-day journey across the solar disk to the west limb. This means that a CNP spotted on the east limb has at least a 51.6\,\% chance of reaching the west limb. This can either mean that all visible CNPs have lifetimes of about 21 days (i.e., half-lives of 14 days), or that a subset of about half of the visible CNPs have much longer lifetimes.\par
 	It is important to note that the location of null points on the Sun's east limb were determined from information obtained after modeling one complete Carrington rotation. However, it is possible that the east limb crossing time corresponded to the previous Carrington rotation, but the magnetogram information from the previous CR would be close to 20 days old when reaching the east limb. So we decided to use the information that would be available only six to seven days after crossing the east limb and then move the feature backwards.  Such a possibility of CNPs crossing temporal boundaries between Carrington rotations should be kept in mind when using the Carrington rotation information given in Table 1.\par
 	This catalog consists of CNPs from a limited resolution PFSS model and therefore represents nulls associated more with the high-altitude interaction between active regions and the global dipole field, than the lower quiet Sun. A finer resolution of the magnetic field around an active region can be obtained, for example, by using a NLFFF extrapolation as done by \citet{Sun12} for AR\,11158 to find more nulls. However, the resolution used here works in our favor by establishing a lower limit to the size of bracketing needed in Greene's three-dimensional bisection method for finding null points. The fields will be well behaved within the 48 Mm uncertainty limit and this will prevent an inaccurate null-point count with Greene's method as discussed by \citet{Haynes07}.

 \begin{center}
\begin{longtable}{cc}
\caption{List of p-values from Kolmogorov--Smirnov testing of observed and predicted CNP parameters. Latitude is the only parameter that shows a significant difference between populations.}
\label{tbl:2}\\
\hline
\bf Parameter & \bf P-Value \\  \hline
Latitude & 0.004 \\ 
Radial Distance & 0.492 \\ 
Tilt of Spine In the Plane of the Sky & 0.944 \\ 
Tilt of Spine Out of the Plane of the Sky & 0.857 \\ 
Ratio of Eigenvalues Associated with Fan & 0.999 \\ 
Eigenvalue Associated with Spine & 0.447 \\ 
\end{longtable} 
\end{center}
 	    
     \subsection{CNP Parameters and Observability}
\begin{figure}
\centering
  \begin{tabular}{@{}cccc@{}}
    \includegraphics[width=.82\textwidth]{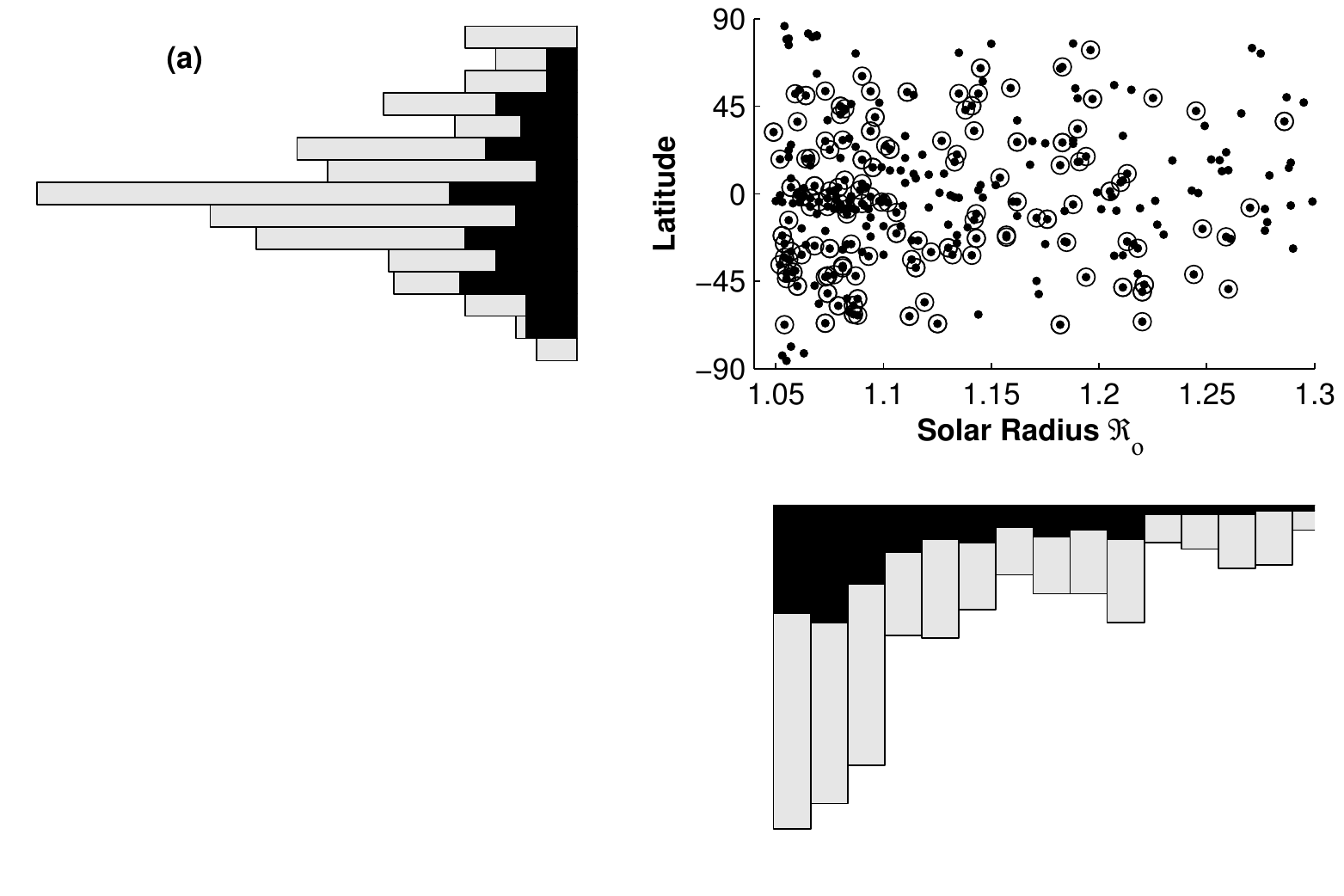} \\
    \includegraphics[width=.82\textwidth]{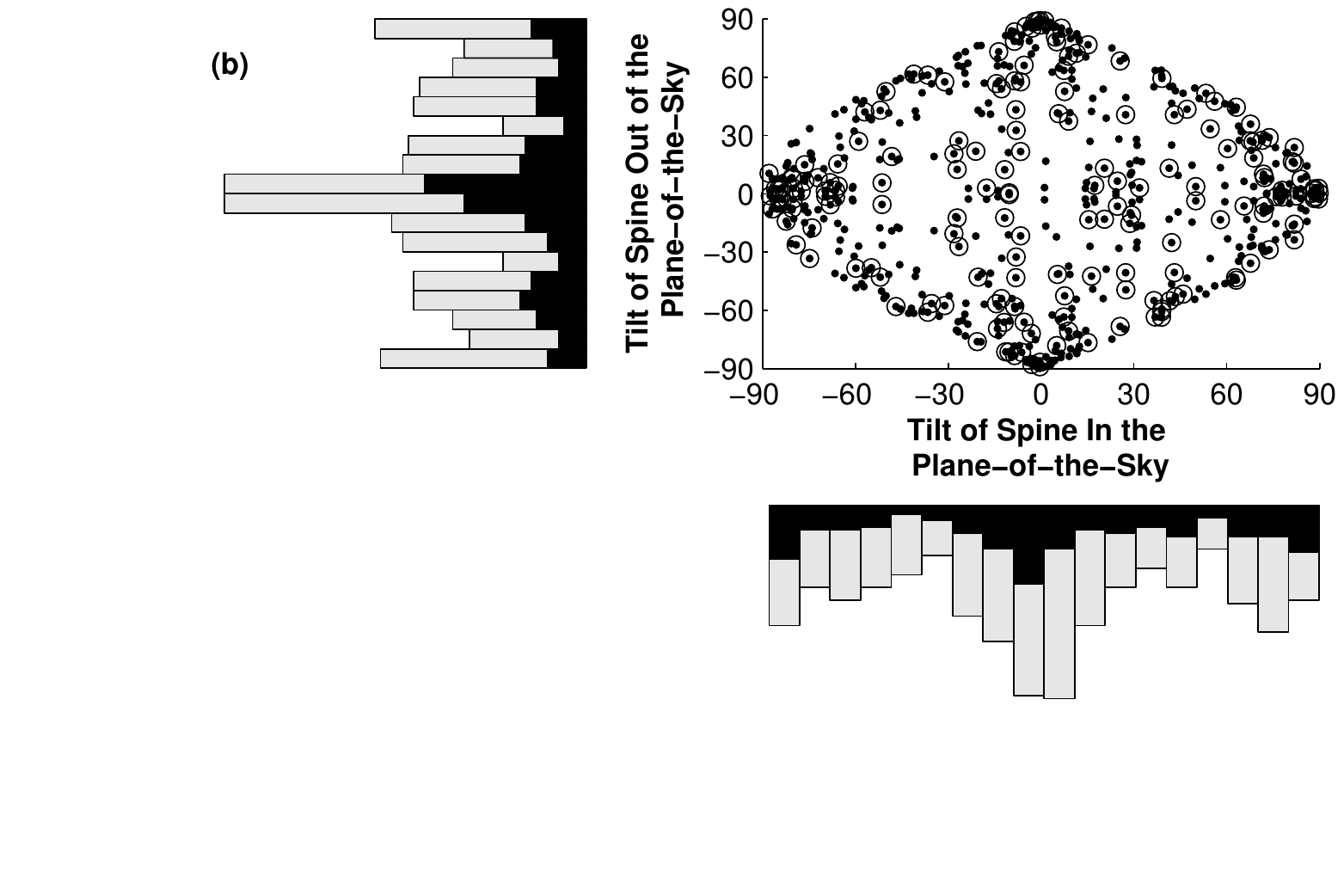} \\
    \includegraphics[width=.82\textwidth]{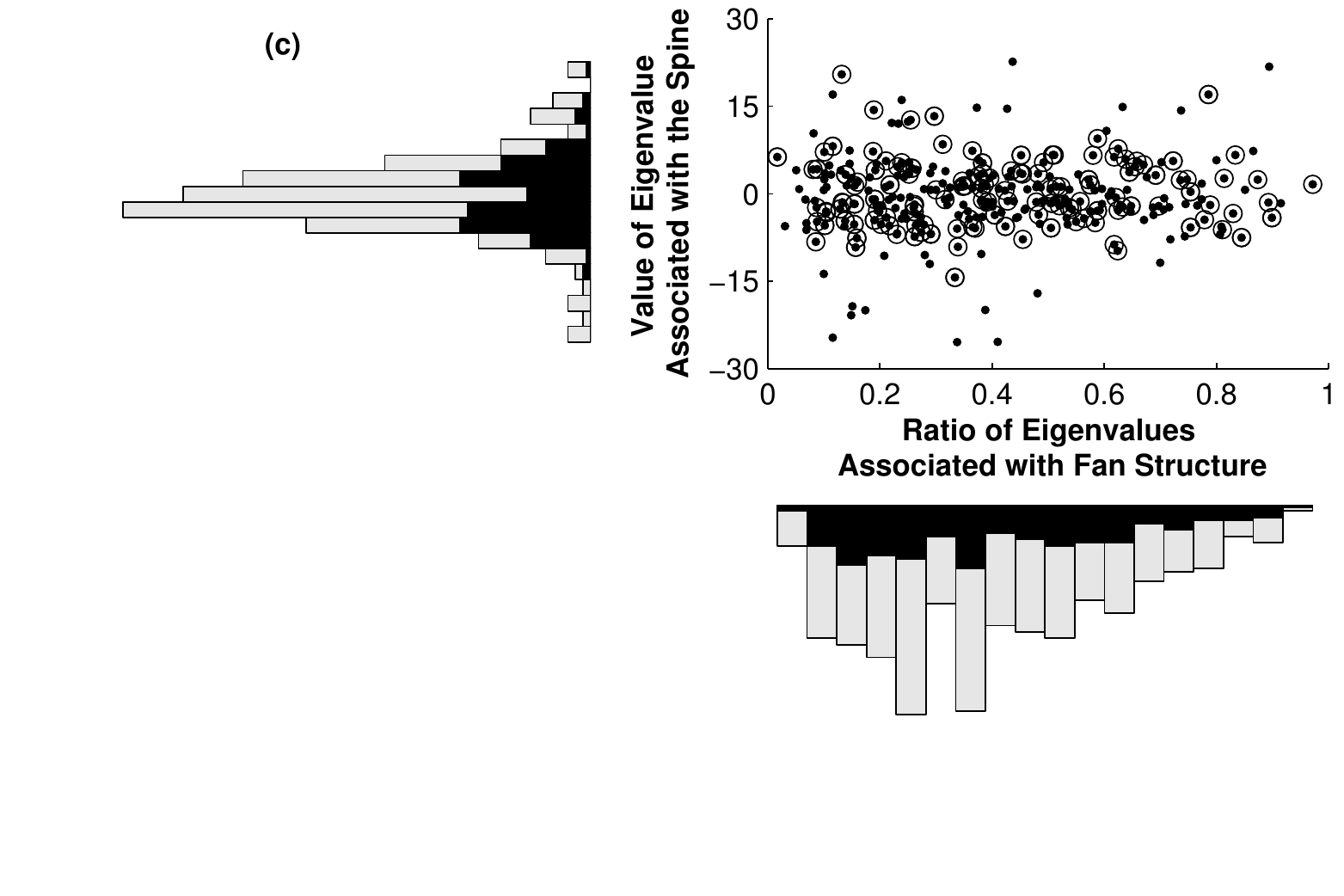} \\
  \end{tabular}
  \caption{Scatter plots showing location of null points on the east and west solar limb for CR 2098 to 2139. Histograms on the margins show the distribution of each quantity for predicted (dots and grey bars) and observed (circles and black bars) results. The latitudinal locations in (a) indicate observed CNPs  differed significantly from predicted ones, which was confirmed by a Kolmogorov--Smirnov test. No significant difference was found, between the predicted and observed populations, for any of the other quantities shown in (a), (b), or (c).}
   \label{distribution_plots}
\end{figure}

     The observed null points were examined further to determine if any of the calculated parameters from the PFSS model had an effect on observability for CR 2098 to 2139. The number of observed null points with a negative sign was found to be 51.1\,\% on the west limb and 45.2\,\% on the east limb. The sign appears to have no significance for the observability of CNPs. Scatter plots with histograms are shown for CNPs' latitude and radial distance (Figure \ref{distribution_plots} a), tilt of spine in and out of the plane of the sky (Figure \ref{distribution_plots} b), and ratio of eigenvalues forming the fan and eigenvalue of spine (Figure \ref{distribution_plots} c). The 582 predicted CNPs are indicated with dots and grey bars in the figures, while the 183 observed CNPs are indicated with circles and black bars. The CNP distributions drop off as a function of radius (Figure \ref{distribution_plots} a and \ref{radial_distribution}) and have a slight bias toward observability in the southern hemisphere from Figure \ref{distribution_plots} a. One possible cause for a observational bias is the Sun's well-known hemispheric asymmetry in activity and \changes{is explored in Section 4.3}. This bias was confirmed by performing a Kolmogorov--Smirnov (K--S) test to determine if the observed and predicted populations came from the same continuous distribution. A K--S p-value of 0.004 was found with respect to the latitude. The p-value is used as an indication for accepting or rejecting the null hypothesis that the distributions are identical. In the case of p-value$\,=\,$0.004, we can state that the two sample populations do not  come from the same parent population at the 99.6\,\% significance level. This, along with the latitudinal histogram distribution shown in Figure \ref{distribution_plots} a, indicates a possible observational bias toward the southern hemisphere during this three-year survey. Similar p-values can be found in Table \ref{tbl:2} for the other five parameters shown in Figure \ref{distribution_plots}. However, none of the other parameters showed significant differences between observed and predicted distributions. The observed CNPs seemed to avoid small eigenvalue ratios associated with the fan structure by inspection of Figure \ref{distribution_plots} c, but there was no indication of this difference from the K--S test. Future work will be necessary to determine whether this double peak is real or just an artifact of the small sample size.\par
     
  \begin{figure} 
\centerline{\includegraphics[width=0.7\textwidth,clip=]{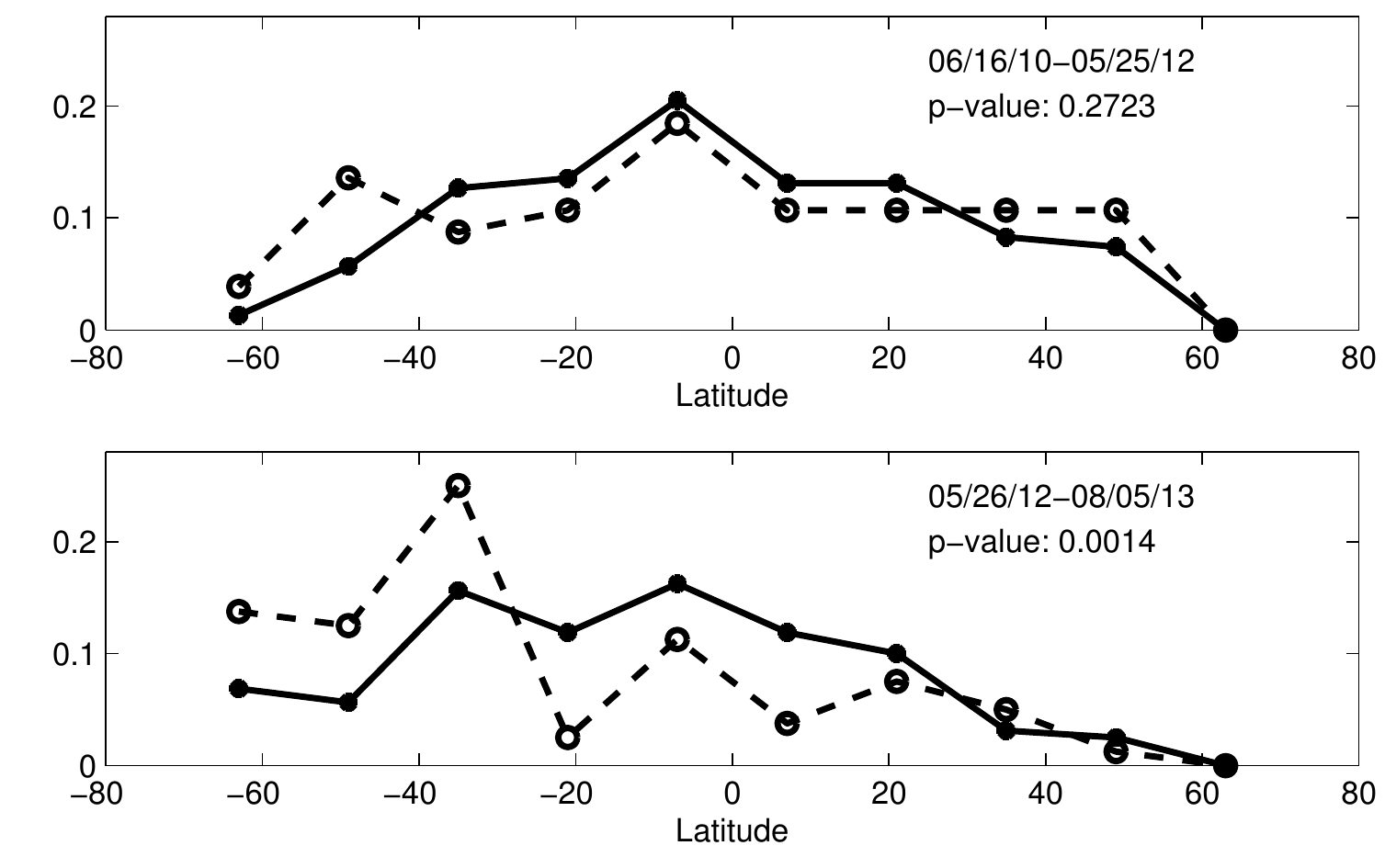}}
 \caption{The latitudinal distribution of CNPs when the northern (top) and southern (bottom) hemispheres were dominant according to sunspot areas \protect\citep{Chowdhury13}. All of the observed and predicted CNPs are indicated by the dashed and solid line, respectively. Each marker represents the normalized count of CNPs with binning applied every 14 degrees.}
 \label{latitude_distribution}
 \end{figure}

          \subsection{Hemisphere Asymmetry}
     The K--S test result for the predicted and observed latitudinal distributions of CNPs is explored further to determine if this was possibly due to a hemispheric asymmetry in solar activity. All of the observed and predicted CNPs were first broken into two different groups as shown in Figure \ref{latitude_distribution}. The top graph consists of 103 observed and 229 predicted CNPs between 16 June 2010 to 25 May 2012, when the solar activity was dominated by the northern hemisphere; also, the bottom graph has 80 observed and 160 predicted CNPs between 26 May 2012 to 15 August 2013, when the southern hemisphere became dominant according to sunspot areas \citep{Chowdhury13}. The dashed line indicates the normalized latitudinal distribution of observed CNP counts, while the solid line shows the same for the predicted CNPs. There was a noticeable increase in observed CNPs from the southern hemisphere as it became more active. Another K--S test was performed for the two sets of latitudinal distribution shown in Figure \ref{latitude_distribution}. The resulting p-value dropped from 0.2723 to 0.0014 as the southern hemisphere became more active. This means the observed and predicted CNPs did not come from the same parent population at the 72.7\,\% and then at the 99.9\,\% significance level. Therefore, the hemispheric asymmetry does have a statistically significant effect on the observed and predicted latitudinal distribution of CNPs.\par
\begin{figure}
 \centerline{\includegraphics[width=01.0\textwidth]{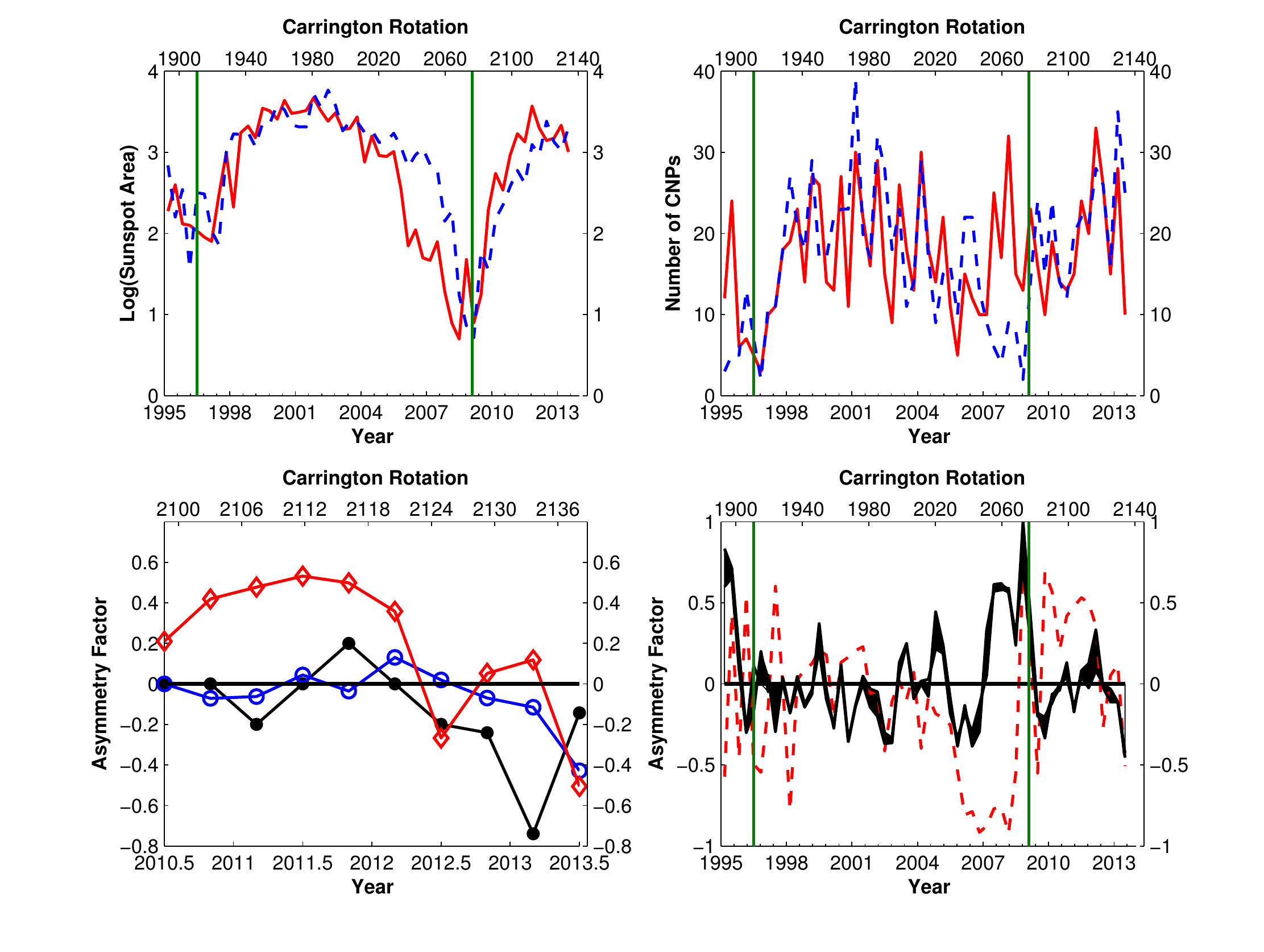}} 
\caption{(a): Sunspot areas with the red solid line denotes the northern hemisphere and the blue-dashed line is for southern hemisphere. (b): Number of \changes{predicted} CNPs in the northern hemisphere are indicated by the red line and southern hemisphere ones are shown as a blue-dashed line. (c): Indicates the asymmetry factor for sunspot areas (red with diamond markers), observed CNPs (black with filled circle markers), and predicted CNPs (blue with open circle markers). The general trend indicates a decrease in the northern hemisphere's strength over the three-year survey. (d): Asymmetry factor for sunspot areas (red-dashed line) and predicted CNPs (black line). The asymmetry factor from CNPs indicated more of a southern hemisphere dominance at the end of Cycle 23, which is in contradiction to results found from examining sunspot areas. The varying thickness of the black line in Figure \ref{asym_plots} d indicates the range of possible {\it A}-values. The thickness was determined by including and excluding the CNPs located within $\pm 6.2$ degrees from the Equator when calculating {\it A}. This accounts for the spatial uncertainty of the PFSS model used. The green vertical lines in Figures 8 a, b, and d indicates the beginning and end of Solar Cycle 23.}
	\label{asym_plots}
\end{figure}

     An additional investigation was conducted to determine whether there was a correlation between CNP counts and the frequently used sunspot area data for indicating hemispheric activity. Figure \ref{asym_plots} a shows the log of sunspot areas (in units of millionths of a hemisphere)  for the northern hemisphere (indicated by the red solid line) and southern hemisphere (indicated by the blue-dashed line) obtained from the USAF/NOAA sunspot-data archive. \changes{The beginning (1996.5) and end (2009.1) of Solar Cycle 23 are indicated by vertical green lines in Figures \ref{asym_plots} a, b, and d \citep{Chowdhury13}.} There was a noticeable dominance of the southern hemisphere at the end of Cycle 23 and then the northern one at the beginning of Cycle 24. The same thing was done for the CNP count in each hemisphere and these results were plotted in Figure \ref{asym_plots} b with the same color scheme. The CNPs indicated more of a southern hemisphere dominance at the end of Cycle 23, which was in contradiction to results found from examining sunspot areas. An additional check was made by using the asymmetry factor [{\it A}] defined as
 \begin{equation}
A=\frac{N-S}{N+S}
\end{equation}        
by \citet{Chowdhury13}. The {\it N} here represents either the sunspot area or CNP count in the northern hemisphere and {\it S} indicates the same for the southern hemisphere. This new parameter was plotted in Figure \ref{asym_plots} c for the sunspot areas (red with diamond markers), observed CNPs (black with filled circle markers), and predicted CNPs (blue with open circle markers). The general trend indicates a decrease in the northern hemisphere's strength over the three-year survey. However, when the asymmetry factor for predicted CNPs (black line), going back to CR 1893, was plotted against sunspot areas (red-dashed line) in Figure \ref{asym_plots} d, a noticeable difference is seen once again near the end of Cycle 23 around 2008. \changes{The sunspot areas, and the observed and predicted CNP counts in Figure \ref{asym_plots} c all show a decrease in their asymmetry factor with respect to time, which indicates the southern hemisphere is becoming more dominant around the maximum of Solar Cycle 24. However, Figures \ref{asym_plots} a and b show that this agreement does not hold up well in general.} Therefore, CNPs do not appear to act reliably as a separate proxy for determining the strength of the hemispheres. All of the data presented in Figure \ref{asym_plots} used a binning of four months. Figure \ref{asym_plots} b and d used predicted CNPs between $1.05\,\rm R_{\odot}<r<2.5\,R_{\odot}$. The thickness of the black line in Figure \ref{asym_plots} d indicates the range of possible {\it A} values, which was determined by including and excluding spatially uncertain CNPs located within $\pm\,6.2$ degrees from the equator when calculating {\it A}. \par

\subsection{Temporal, Radial, and Latitude Distribution of CNPs}
\begin{figure}
 \centerline{\includegraphics[width=0.7\textwidth]{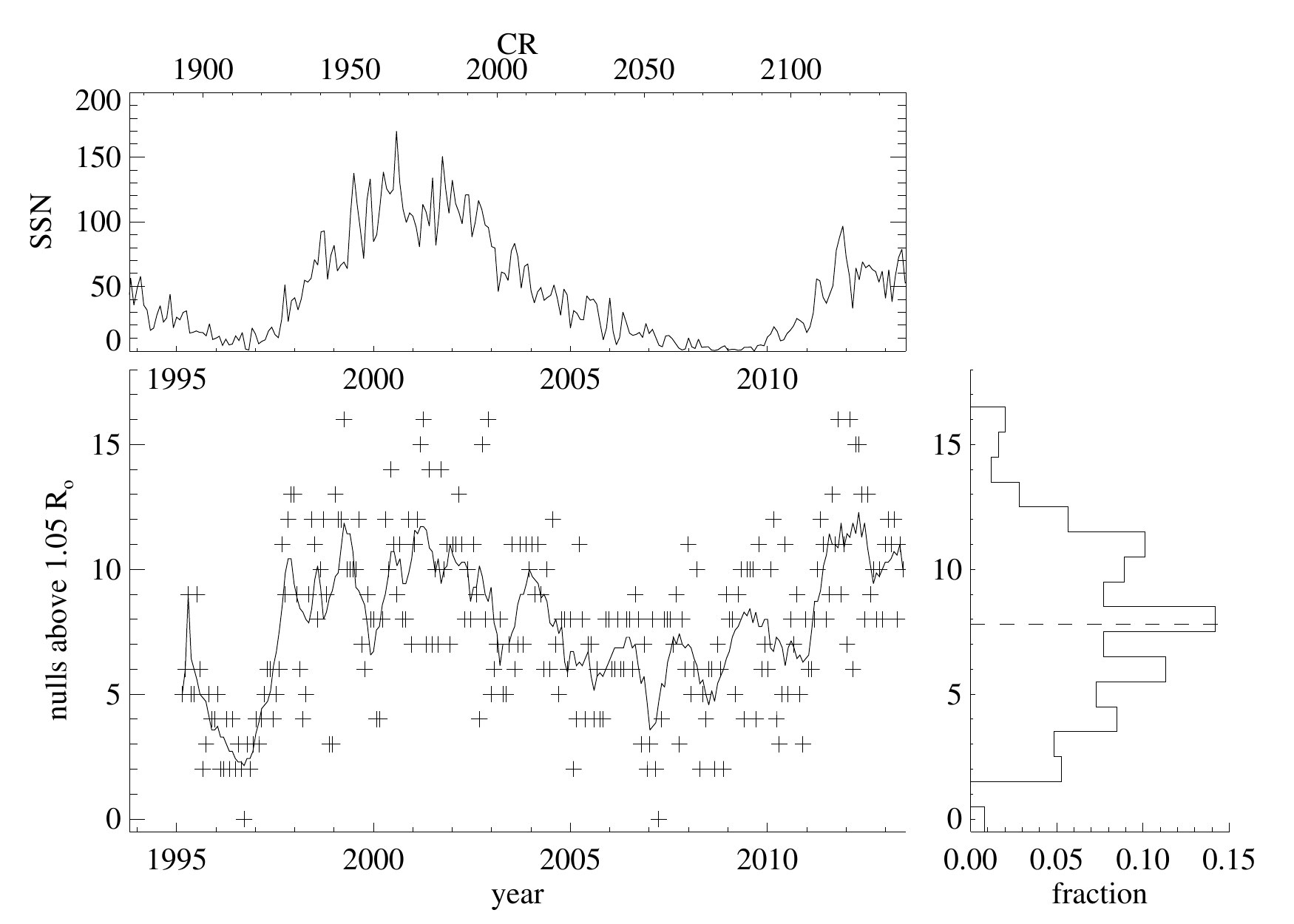}} 
\caption{The distribution of predicted null points found in the PFSS fields {\em vs.} time.  Plusses in the central panel show the number of null points in each rotation; the solid curve is a smoothed version.  The right panel shows a histogram of the numbers in each rotation with the mean indicated with a horizontal dashed line.  The International Sunspot Number is plotted in the top panel.}
	\label{temporal_distribution}
\end{figure}

Applying the technique described in Section 2.1 to the 247 Carrington rotations from 1893 to 2139 
\changes{(March 1995 to August 2013)}, we found 1924 distinct null points within the range 
$1.05\,\rm R_{\odot}\le r < 2.5\,R_{\odot}$: a mean of 7.8 nulls within any rotation.  As shown in Figure\ \ref{temporal_distribution},
the number of null points track the activity cycle, ranging up to 16 around solar maximum and down to zero or two during solar minimum. \citet{Platten14} found the opposite trend by using higher-resolution {\it Kitt Peak} and {\it  Synoptic Optical Long-term Investigations of the Sun} (SOLIS) magnetogram data to incorporate spherical harmonics up to $\ell=81$ into the PFSS model. This resulted in the identification of nulls below $r=1.05\,\rm R_{\odot}$ that are associated with the quiet Sun, which will increase in number at solar minimum. However, Figure 19 of \citet{Platten14}  shows that the number of nulls located above $r=1.05\,\rm R_{\odot}$ agrees with the lower resolution results found here.\par
\begin{figure}
 \centerline{\includegraphics[width=0.7\textwidth]{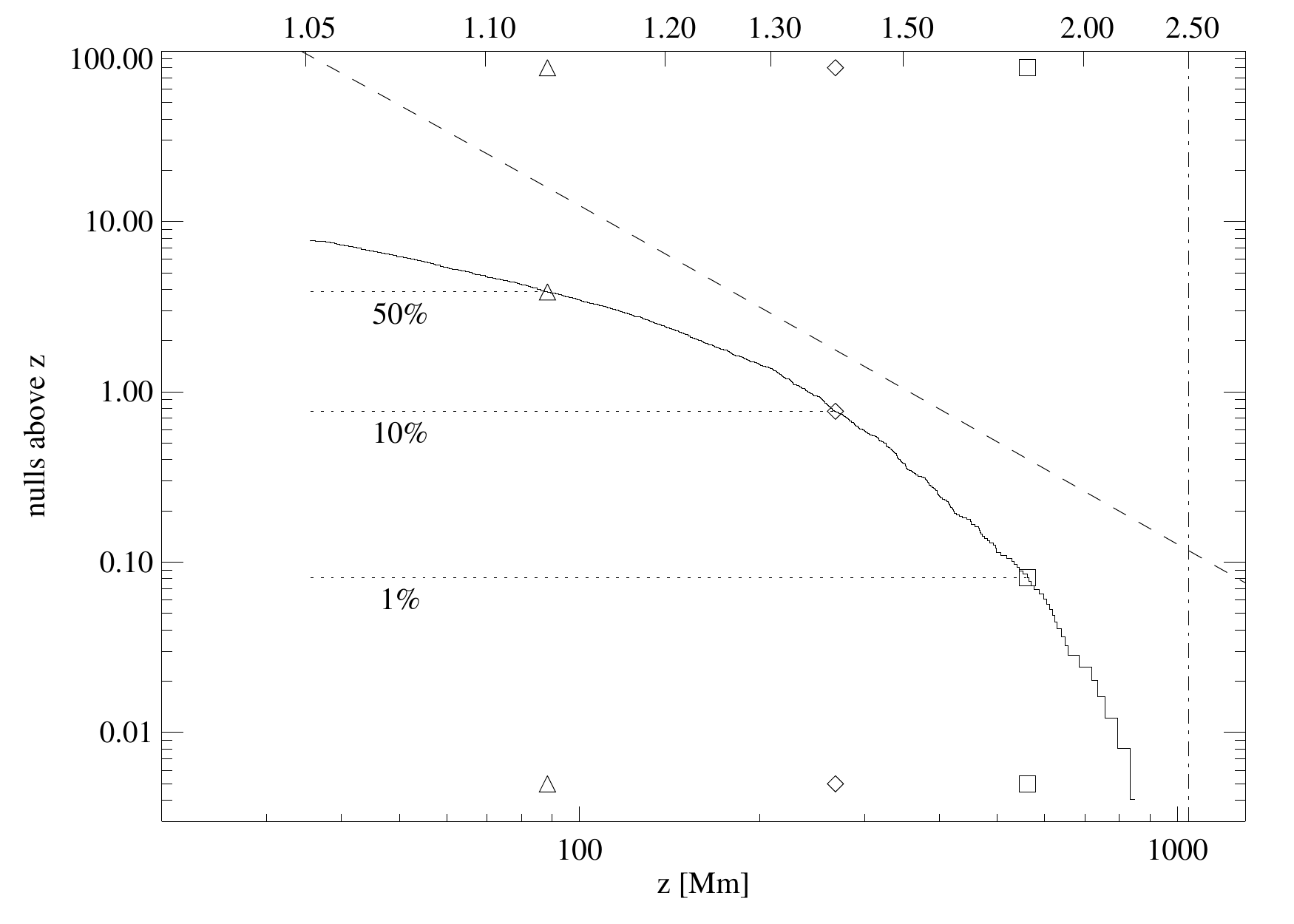}} 
\caption{The distribution of null points found in the PFSS fields {\em vs.} height.  The solid curve shows the average number of null points, in a single Carrington rotation, above a given height.  The lower axis shows height in Mm above the solar surface, while the upper axis shows the height in solar radii.  The dashed curve shows the distribution found by \protect\citet{Longcope09}.}
	\label{radial_distribution}
 \end{figure}

The distribution in height of all null points (irrespective of time), Figure\ \ref{radial_distribution}, shows that 90\,\% of the null points are below $r=1.38\, \rm R_{\odot}$.  The number found comes within a factor of two of the null column density
\begin{equation}
  N(z) ~=~ {0.021\over (z + 1.5\,{\rm Mm})^2 } ~~,
\end{equation}
that \citet{Longcope09} found in a statistical analysis of quiet-Sun magnetic fields over the same period.  Even this level of agreement is remarkable due to the very different natures of the two fields: PFSS structured on large scales {\em vs.} quiet-Sun field assumed to be statistically homogeneous.  The solid curve in Figure\ \ref{radial_distribution} rolls over at $r=R_{\rm SS}=2.5\,\rm R_{\odot}$.  It also rolls over at low heights due to the limited resolution of a PFSS limited to $\ell\le29$.\par
\begin{figure}
 \centerline{\includegraphics[width=0.7\textwidth]{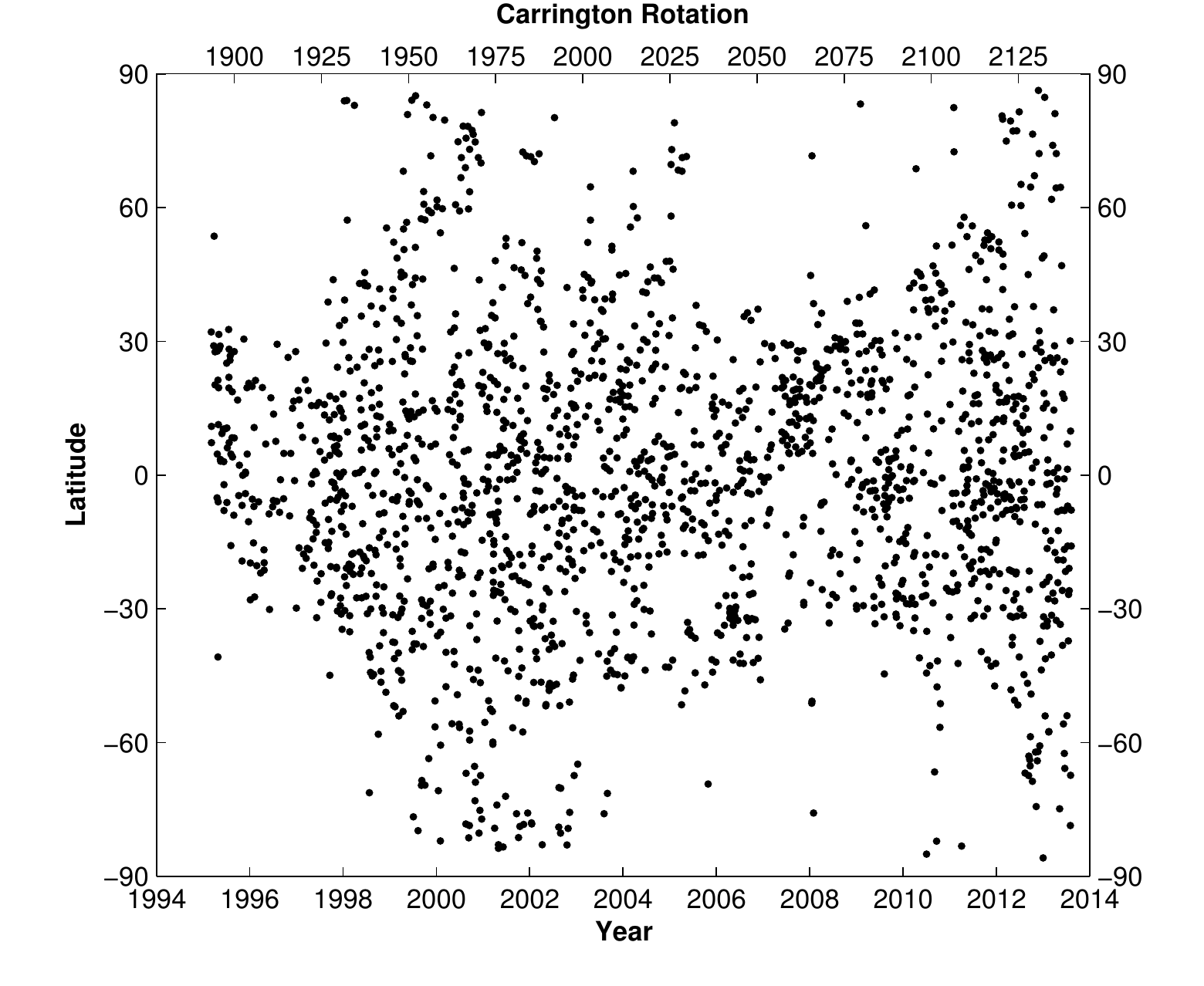}} 
\caption{The latitudinal distribution of predicted null points found in the PFSS fields {\em vs.} time. There is a consistent band of null point creation between $\pm$30$^{\circ}$ in latitude; higher latitudes see activity at solar maximum.}
	\label{lat_time_dist}
\end{figure}

All of the predicted CNPs' latitudinal positions were plotted as a function of time in Figure \ref{lat_time_dist}. This shows that CNPs always exist near the Equator regardless of the solar-cycle stage. Our finding differs from that of \citet{Cook09}, which showed that the distribution would resemble the familiar butterfly diagram observed with sunspot areas. CNPs can be found at higher latitudes during solar maximum, but there are no corresponding gaps near the Equator during this time, in agreement with the findings of \citet{Platten14}.

\section{Conclusions}	
\changes{A multi-year catalog of observations associated with coronal null point structures has been compiled for the first time.} This was possible because of the continuous full-limb and high-cadence coverage of the Sun from SDO/AIA, in conjunction with pre-existing PFSS modeling. A radial-gradient filter of  the 171\,\AA\ band was then incorporated to manually inspect 582 predicted null-point sites on the east and west limbs.\par
Our method shows the theoretical PFSS model can be used to indicate the location of CNPs in SDO/AIA 171\,\AA\ images, with an accuracy of approximately 31\,\%. However, this is a lower limit since images of 18.6\,\% of the predicted CNP sites were saturated. The predicted crossing times were adequate for determining the presence of CNPs, with an error in classification made \changes{6.7\,\% of the time. (This was determined from the total of 39 out of the 582 CNPs listed in the catalog that were incorrectly classified by first inspection, before they were examined further with {\sf JHelioViewer}.)} This can be useful for estimating the number of CNPs missed when users inspect SDO/AIA images only at the predicted limb-crossing time. Over half of the observed CNPs on the east limb were also stable enough to be seen with structure upon reaching the west limb. CNPs appear continuously throughout the solar cycle near the Equator. There are no gaps near solar minimum, which prevents them from appearing like a butterfly diagram as is commonly seen with sunspots. CNPs do not act as a useable proxy for determining which hemisphere of the Sun is dominant. However, the predicted nulls located at $r\ge1.05\,\rm R_{\odot}$ can be used to indicate when there is an increase in solar activity. The radial distribution of predicted CNPs is within a factor of two from the results indicated by \citet{Longcope09} for quiet Sun. There is also a noticeable difference in the latitudinal distribution of predicted and observed CNPs due in some part to the asymmetrical hemisphere activity. Surprisingly, there appears to be no effect on observability of CNPs from spine orientation. An image catalog and comprehensive list like Table \ref{tbl:1} is available on-line for all 582 CNPs.\par

\changes{This work used low resolution ($\ell=29$) magnetogram data from WSO as input to the PFSS model. \citet{Platten14} showed how using higher resolution ($\ell=81$) magnetograms to construct the magnetic-field topology can increase the number of predicted CNPs. However, increasing the number of spherical harmonics does not negate the number of CNPs reported here, it will only find more CNPs at a higher accuracy.}\par

Further investigation is necessary to determine whether there is actually a bias toward not observing CNPs with low eigenvalue ratios associated with the fan. \changes{The intent of future work is to also explore the separatrix surfaces associated with the observed CNPs found in our survey.} This work can also be useful in future efforts to mine AIA data for additional CNPs, as shown by \citet{Martens12}, and to investigate their temporal evolution. There are still many questions that need to be addressed: Why do some of the null points remain stable for so long while others quietly disappear? What causes some of the CNPs to erupt into solar flares? This work does not address these questions, but it is believed that further work on the complexity of the surrounding active regions will shed some light on these matters. \par

%

%
\begin{acks}
We would like to thank Montana Space Grant Consortium for a fellowship to conduct this work. This work was also partially supported by NASA under contract SP02H3901R from Lockheed-Martin to Montana State University. The authors like to thank Spiro Antiochos, and Sophie Masson for their useful discussions, and the helpful feedback of the anonymous referee.
\end{acks}

%

 \bibliographystyle{spr-mp-sola}
 \bibliography{SOLA_D_14_00144_bibliography}   
%
%
%
%
\end{article} 
\end{document}